\documentclass[aps,prb,twocolumn,showpacs,preprintnumbers,superscriptaddress,floatfix]{revtex4-1}
\usepackage{graphicx}
\usepackage{dcolumn}
\usepackage{float}
\usepackage{bm}
\usepackage{latexsym}
\usepackage{array}
\usepackage{amssymb}
\usepackage{amsfonts}
\usepackage{amsmath}
\usepackage{color}

\newcommand{\done}[1]{}

\newcommand{\fref}{f_{\rm ref}}
\newcommand{\rhoref}{\rho^{\rm ref}}

\newcommand{\sqbra}[1]{\left[#1\right]}
\newcommand{\cubra}[1]{\left\{#1\right\}}
\newcommand{\robra}[1]{\left(#1\right)}

\newcommand{\de}[1]{d{#1}\,}

\newcommand{\be}{\begin{equation}}
\newcommand{\ee}{\end{equation}}
\newcommand{\bea}{\begin{eqnarray}}
\newcommand{\eea}{\end{eqnarray}}
\newcommand{\baa}{\begin{align}}
\newcommand{\eaa}{\end{align}}
\newcommand{\br}{{\bm r}}

\newcommand{\brp}{{\bm r'}}

\newcommand{\sigmap}{{\sigma_1}}

\newcommand{\eqn}[1]{Eq.~\eqref{#1}}
\newcommand{\fig}[1]{Fig.~\ref{#1}}

\newcommand{\sect}[1]{Sec.~\ref{#1}}
\newcommand{\appendixx}[1]{Appendix~\ref{#1}}

\newcommand{\dbracket}[2]{\langle #1|#2\rangle}
\newcommand{\expect}[2]{\langle {#1|#2|#1}\rangle}

\newcommand{\nalsi}{n_{\alpha\sigma}}

\newcommand{\falsi}{f_{\alpha\sigma}}

\newcommand{\Vext}{{V_{\rm ext}}}

\newcommand{\Ex}{E_{\rm x}}
\newcommand{\EH}{E_{\rm H}}

\newcommand{\Exc}{E_{\rm xc}}
\newcommand{\EHxc}{E_{\rm Hxc}}

\newcommand{\vxc}{v_{\rm xc}}
\newcommand{\vHxc}{{v_{\rm Hxc}}}
\newcommand{\kHxc}{h_{\rm Hxc}}

\newcommand{\inbra}[1]{{#1}}

\newcommand{\etal}{{\it et al.}}

\usepackage{ifthen}
\usepackage{calc}
\usepackage{url}
\usepackage{fancyvrb}
\usepackage[utf8]{inputenc}
\usepackage[english]{babel}%
\usepackage{mathrsfs}
\newcommand{\mhs}[1]{\mathscr{#1}}
\begin{document}

\title{{Koopmans-compliant functionals and their performance 
against reference molecular data}}
\author{Giovanni Borghi}
\email{giovanni.borghi@epfl.ch}
\affiliation{Theory and Simulations of Materials (THEOS), and National Center for Computational Design and Discovery of Novel Materials (MARVEL), \'Ecole Polytechnique F\'ed\'erale de Lausanne,
1015 Lausanne, Switzerland}
\author{Andrea Ferretti}
\affiliation{Centro S3, CNR--Istituto Nanoscienze, I-41125 Modena, Italy}
\author{Ngoc Linh Nguyen}
\affiliation{Theory and Simulations of Materials (THEOS), and National Center for Computational Design and Discovery of Novel Materials (MARVEL), \'Ecole Polytechnique F\'ed\'erale de Lausanne,
1015 Lausanne, Switzerland}
\author{Ismaila Dabo}
\affiliation{Department of Materials Science and Engineering, Materials Research Institute, and Penn State Institutes of Energy and the Environment, The Pennsylvania State University, University Park, PA 16802, USA}
\author{Nicola \surname{Marzari}}
\affiliation{Theory and Simulations of Materials (THEOS), and National Center for Computational Design and Discovery of Novel Materials (MARVEL), \'Ecole Polytechnique F\'ed\'erale de Lausanne,
1015 Lausanne, Switzerland}
\begin{abstract}
Koopmans-compliant functionals emerge naturally from 
extending the constraint of piecewise linearity 
of the total energy as a function of the number
of electrons to each fractional orbital occupation.
When applied to approximate density-functional
theory, these corrections give rise to orbital-density-dependent 
functionals and potentials. We show that the simplest implementations of Koopmans' compliance
provide accurate estimates for the quasiparticle excitations
and leave the total energy functional almost or exactly intact, i.e., they describe correctly electron removals or additions, but do not necessarily alter the electronic charge density distribution within the system. Additional functionals can then be constructed that modify the potential energy surface, including e.g. Perdew-Zunger corrections. These functionals become exactly one-electron self-interaction free and, as all Koopmans-compliant functionals, are approximately many-electron self-interaction free.
We discuss in detail these different formulations, and 
provide extensive benchmarks for the 55 molecules in the reference G2-1 set, using
Koopmans-compliant functionals constructed from
local-density or generalized-gradient approximations.
In all cases we find excellent performance in the electronic properties, comparable or improved with respect to that of many-body perturbation theories, such as G$_0$W$_0$ and self-consistent GW, at a fraction of the cost and in a variational framework that
also delivers energy derivatives. Structural properties and
atomization energies preserve or slightly improve the accuracy of the underlying density-functional approximations (Note: Supplemental Material is included in the source).
\end{abstract}

\maketitle

\section{Introduction} 
A key advantage of Kohn-Sham (KS) density-functional theory
(DFT)~\cite{Hohenberg1964, Kohn1965} over wave-function approaches is
its combination of accuracy and relatively affordable computational
costs, stemming from the diagonalization of an effective
non-interacting KS Hamiltonian.
DFT is a theory of total energies, and so eigenvalues of the effective KS Hamiltonian have no formal justification as quasiparticle excitations:
Kohn-Sham electrons are auxiliary particles whose wave functions provide 
a parametrization of the total density of the system and a well-defined decomposition of the total energy functional. Notwithstanding this limitation,
it can be proved that the highest-occupied molecular orbital (HOMO) of exact KS-DFT
is equal to the negative of the first ionization energy,\cite{perd-levy83prl,perd-levy97prb} 
since the ionization energy
determines the decay of the charge density in vacuum, and exact DFT 
reproduces this exactly (see also Ref.~[\onlinecite{Chong2002}] concerning the accuracy of KS eigenvalues in approximating charged excitation energies).
On the other hand, common approximations [such
as the local-density (LDA)~\cite{Perdew1981} or 
generalized-gradient PBE~\cite{Perdew1996}] provide
HOMO eigenvalues that display large discrepancies from the exact
values.

Such failures have been 
connected~\cite{cohe+08sci,Dabo2009,Dabo2010,krai-kron13prl} to
the deviation from piecewise linearity~\cite{perd+82prl} (PWL) of the total
energy as a function of particle number, and the associated derivative
discontinuity at integer numbers. In approximate functionals, the total energy is usually continuous and
convex, with a discontinuity in the first derivative which lacks the contribution from the exchange-correlation potential~\cite{Sham1983}; as a result of the convexity, the HOMO eigenvalue is too high in energy and
the ionization energy is under-estimated.

The importance of piecewise linearity to improve approximate energy functionals was actually first
discussed in the context of Hubbard corrections to DFT~\cite{coco-degi05prb,kuli+06prl,cococcioni_thesis},
where DFT+$U$ had been viewed as restoring PWL for a localized Hubbard manifold in contact
with the reservoir of an extended solid~\cite{coco-degi05prb}.
This point of
view was extended to strongly localized transition-metal centers in molecules~\cite{kuli+06prl},
arguing that piecewise linearity was actually correcting strong self-interactions, and thus was meaningful even in the
single-site limit.  Moreover, deviation from PWL has been
suggested~\cite{ruzs+06jcp,mori+06jcp,cohe+08sci} as a
definition of electronic self-interaction errors (SIE's) in the context of
many-electron systems.

PWL is then recognized as one of the most relevant features to address in order
to improve on the accuracy of approximate functionals. Both the Koopmans-compliant functionals introduced by Dabo \etal~\cite{Dabo2009,Dabo2010,Dabo2013} and a number of other approaches have been proposed to correct for the missing PWL~\cite{lany-zung10prb,zhen+11prl,refa+11prb,refa+12prl,krai-kron13prl,kronik_hybrid}
(notably, in recent work Kraisler and Kronik~\cite{krai-kron13prl} implement a correction which is 
very similar to the K Koopmans' correction~\cite{Dabo2009,Dabo2010} and formally identical to what is called KI in this work,
but restricted to the frontier orbitals alone).

In this paper we focus on imposing Koopmans' compliance to
approximate functionals, as discussed earlier,\cite{Dabo2009,Dabo2010,Dabo2013,psik_koopmans}
to introduce PWL conditions that do not rely on pre-defined Hubbard manifolds.
In a nutshell, an approximate functional is made Koopmans-compliant (KC) by removing, orbital-by-orbital, the change in total energy as a function of the fractional occupation of that
orbital, a well-defined Slater integral, usually approximately quadratic, and substituting it with a linear term that
is directly proportional to that occupation. The approach is completely determined
once the linear coefficient is chosen, typically as the slope that best approximates the
exact one. This linear coefficient can be chosen either with a Slater-1/2 approach (i.e., by taking the
orbital energy at 1/2 occupation), as in Ref.~\onlinecite{Dabo2010}, or by taking the difference between energies at the two surrounding integer 
occupations.  These two approaches are almost identical, and provide
a framework and a functional formulation for Slater's original intuition~\cite{Slater1974}, but 
the latter approach (integer Koopmans, or ``KI'', to differentiate it from
the previous approach, labeled here ``K'') is not only more straightforward in its implementation, but provides a deeper insight into the KC formulation, since it can be shown (see Sec.~\ref{Sec:K-KI}) that it preserves exactly the total energy and the wave functions of the underlying approximate functional,
while providing orbital-density-dependent (ODD) potentials that correctly align the
expectation values of the orbitals.
This issue is discussed in detail here and linked to the emergence of scalar contributions (i.e., constant shifts) to the orbital potentials. We show that such scalar potentials originate from the specific functional dependence chosen for the linear slope correction of K or KI.
Improved formulations are also proposed and their accuracy benchmarked.

The paper is organized as follows: in the first part (\sect{Sec:KCdef}) we summarize the formulation
of KC functionals, introduce the different flavors and 
their key features, reporting all explicit expressions for energies and potentials
in the Appendix. In the second part (\sect{Sec:results}) we provide extensive and detailed validation tests, comparing the results for the ionization energies of all molecules in the G2-1 set 
against experimental data or recent many-body perturbation theory
results.~\cite{Rostgaard2010}
Last, we discuss molecular geometries and atomization energies, showing that KC functionals (applied on top of LDA and PBE) not only perform well in the estimation of electronic removal energies, but either preserve the performance of the original functionals in the predictions for these quantities, or improve on them. Technical details related to numerical simulations and the implementation of KC functionals are included in \sect{Sec:technical}, as well as in \appendixx{App:KC}.

\vspace{0.5cm}
\section{{Koopmans-compliant functionals}}\label{Sec:KCdef}
In this section we derive the expressions for the different flavors of Koopmans-compliant corrections. For simplicity, we provide expressions for these corrections when they are applied on top of the LDA functional. Their application on top of PBE, or any other local or semilocal functional, follows straightforwardly. In the following we will use the expression ``base functional'' to refer to the KS functional on top of which the ODD correction is applied.
We start by writing the total density of the system as:
\begin{align}\label{Eq:density_parametrization}
    \rho(\br) = \sum_{i} f_i |\phi_i(\br)|^2, 
\end{align}
i.e., we assume the one-body reduced density matrix of the KS system to be diagonal in the basis $\{\phi_i(\br)\}$, $i$ being a spin-orbital index running over some complete set of orthonormal orbitals.
In this paper we will discuss the case of insulating systems at zero temperature, where $f_i = 1$ for every $i$ labeling a filled orbital, and zero otherwise. In this specific case the index $i$ in~\eqn{Eq:density_parametrization} can be made to run over the $N$ filled orbitals only,
and any unitary mapping within the Hilbert space spanned by $\{\phi_i(\br)\}$ shall leave the total density unchanged.

Let us for the moment consider $f_i$ as external parameters lying between 0 and 1.
We can minimize the LDA energy functional, composed of a kinetic contribution, a Hartree contribution, an exchange-correlation and an external potential term,
%
\begin{widetext}
\begin{align}\label{Eq:DFT_fract_def}
&E^{\inbra{{\rm LDA}}}{[\{f\},\{\phi_i(\br)\}]} = \sum_i f_i \expect{\phi_i}{\hat{T}}+\int \Vext(\br) \rho(\br) \de{\br}
      + \EH\sqbra{\rho}+\Exc\sqbra{\rho} 
     -\sum_{jk} \Lambda_{jk} \robra{\dbracket{\phi_j}{\phi_k}-\delta_{jk}}\,,
\end{align}
\end{widetext}
with respect to $\rho(\br)$ by finding first its minimum with respect to all spin-orbitals $\phi_i(\br)$ 
(subject to the orthonormality constraint $\dbracket{\phi_i}{\phi_j}=\delta_{ij}$).
This yields a total energy $E^{\inbra{{\rm LDA}}}{(\{f\})}$, 
\begin{align}\label{Eq:DFT_fract_def2}
&E^{\inbra{{\rm LDA}}}{(\{f\})} = \min_{\{\phi_i(\br)\}} E^{\inbra{{\rm LDA}}}{[\{f\},\{\phi_i(\br)\}]}\,,
\end{align}
which is function of the occupations $\{f\}$,  summing up to the total number of electrons $N$.
%
%
The LDA total energy, Kohn-Sham eigenvalues $\{\epsilon_i\}$, and eigenvectors $\{\psi_i(\br)\}$ can then be obtained at the end of the minimization over the $\{f\}$ from the eigenvalues and eigenvectors of the matrix of Lagrange multipliers $\Lambda_{ij}$. 
%
As already mentioned, in this paper we will consider only systems in which in the ground state a finite gap separates occupied and unoccupied states, and which have therefore occupation numbers $(\{f\})$ that are either zero or one whenever the total number of electrons $N$ is an integer. For a more general treatment of DFT for systems with a degenerate HOMO and non-pure-state $v$-representable densities, i.e., ground-state densities obtained from KS systems with fractional occupations, we refer for instance to Ref.~[\onlinecite{Kraisler_Makov}].
In the next section, we inspect how the total energy $E^{\inbra{{\rm LDA}}}{(\{f\})}$ changes with the total number of particles $N$, and with the set of occupations $\{f\}$ chosen in the parametrization of Eq.~\eqref{Eq:density_parametrization}.
This analysis will ultimately lead us to the definition of screened Koopmans-compliant energy functionals
\begin{widetext}
\begin{align}\label{Eq:Koopmans_introduction}
&E^{\inbra{{\rm KC}}}{[\{f\},\{\phi_i(\br)\}]} = E^{\inbra{{\rm LDA}}}{[\{f\},\{\phi_i(\br)\}]} +\alpha \sum_i\cubra{f_i\bar\eta_i- \int_0^{f_i} \expect{\phi_i}{\hat{H}_{\rm LDA}(s)}\de{s}}\,,
\end{align}
\end{widetext}
in which orbital-dependent terms are added to the LDA energy in order to restore its missing piecewise linearity, with linear slopes $\bar\eta_i$ whose exact form will be clarified later, and with a screening coefficient $\alpha$ that accounts for orbital relaxations (more generally, 
each orbital should have its own screening coefficient, although in this work we use the same coefficient for all orbitals). The expression $\hat{H}_{\rm LDA}(s)$ stands for the LDA KS Hamiltonian calculated on a charge density where orbital $i$ has an occupation $s$.
\subsection{Changing the number of electrons}
We now discuss what happens if we start from the case of $N_0$ electrons ($N_0$ being integer), in the zero-temperature limit, and we decrease the number of particles by a fractional amount. This is equivalent to computing the energy 
$E^{\inbra{{\rm LDA}}}{(\{f\})}$ in \eqn{Eq:DFT_fract_def2} with $f_{N_0}<1$, leaving all other $f_i$'s equal to one for $i<N_0$, and zero for $i>N_0$. We are in this way simulating the removal of a fractional charge from a system. In the case of many electrons, this procedure requires the fractional charge to be removed from the highest-occupied eigenstate of the Kohn-Sham system (which coincides with a Kohn-Sham orbital); this is also what automatically happens when minimizing the functional {[Eq.~\eqref{Eq:DFT_fract_def2}]}, since the fractionally occupied orbital will end up coinciding with the highest-occupied Kohn-Sham orbital. This fact is nothing but the Aufbau principle for Kohn-Sham DFT, which was proved by Janak~\cite{Janak1978}, and later discussed by other authors~\cite{Giesbertz_Baerends}. 
Also, based on Janak's theorem, the following chain of equalities holds:
\begin{align}\label{Eq:Janak}
\frac{d E^{\inbra{{\rm LDA}}}(N)}{d N}\Big|_{N=N_0^-} = \frac{\partial E^{\inbra{{\rm LDA}}}(N)}{\partial N}\Big|_{N=N_0^-} = \epsilon_{\rm HO}\,,
\end{align}
linking the eigenvalue $\epsilon_{\rm HO}$ of the highest-occupied molecular orbital to the changes of total energy as a function of the number of particles.
The partial derivative refers to changes of the energy by freezing the orbitals to those obtained at $N=N_0$.
The first equality in \eqn{Eq:Janak} is Janak's most important result, which connects the response of the physical system (upon a change in the number of particles) to the response of the fictitious Kohn-Sham system. In the language of the occupation-dependent energy \eqn{Eq:DFT_fract_def2}, and by virtue of the aufbau principle~\cite{Janak1978, perd+82prl, Perdew1981} we can rewrite \eqn{Eq:Janak} as
\begin{align}\label{Eq:Janak1}
\frac{d E^{\inbra{{\rm LDA}}}(\{f\})}{d f_{\rm HO}}\Big|_{f_{\rm HO}=1^-} = \frac{\partial E^{\inbra{{\rm LDA}}}(\{f\})}{\partial f_{\rm HO}}\Big|_{f_{\rm HO}=1^-} = \epsilon_{\rm HO}\,,
\end{align}
where we specify only the value $f_{\rm HO}$ of the occupation of the highest-occupied molecular orbital, the only one involved in the variation (all other occupied orbitals having $f_i=1$). The partial derivative in \eqn{Eq:Janak1} should be computed from the variation of the energy with respect to occupations at frozen orbitals [i.e.,the variation of $E^{\rm LDA}$ in \eqn{Eq:DFT_fract_def2} with orbitals fixed to the value minimizing $E^{\rm LDA}$ for $f_{\rm HO}=1$].

\begin{figure}
\begin{minipage}{\linewidth}
\includegraphics[width=.8\textwidth]{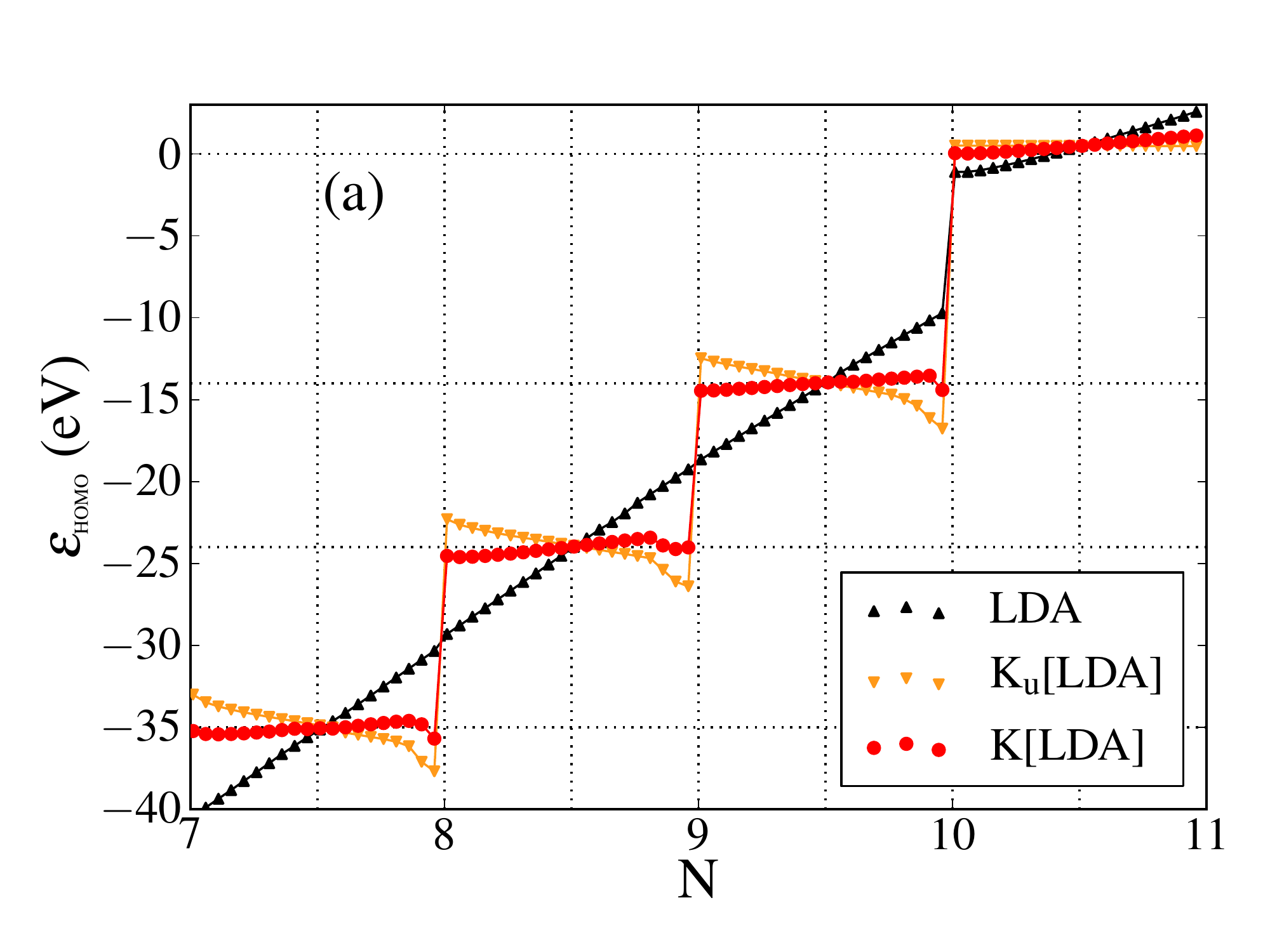}
\includegraphics[width=.8\textwidth]{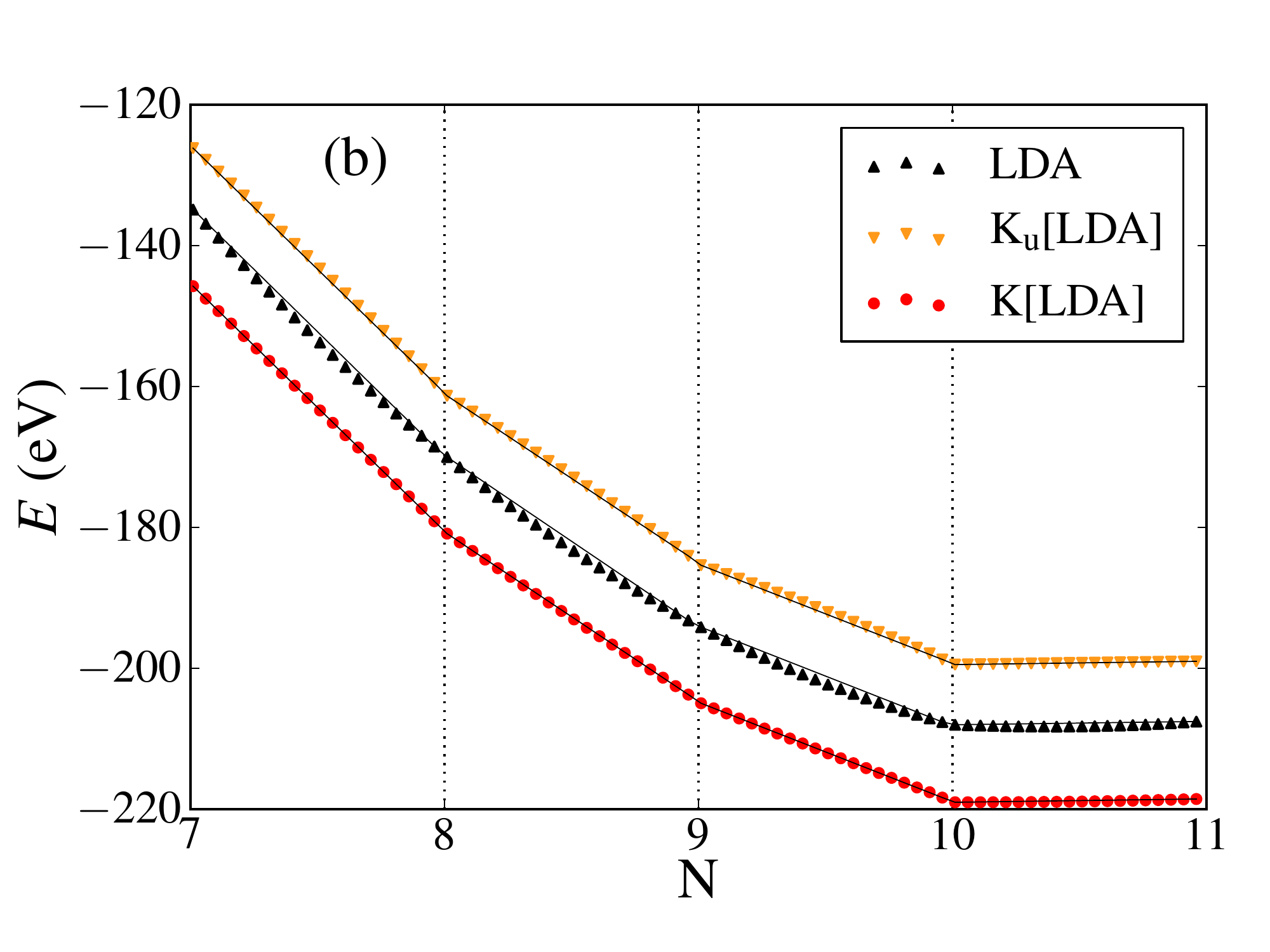}
\includegraphics[width=.8\textwidth]{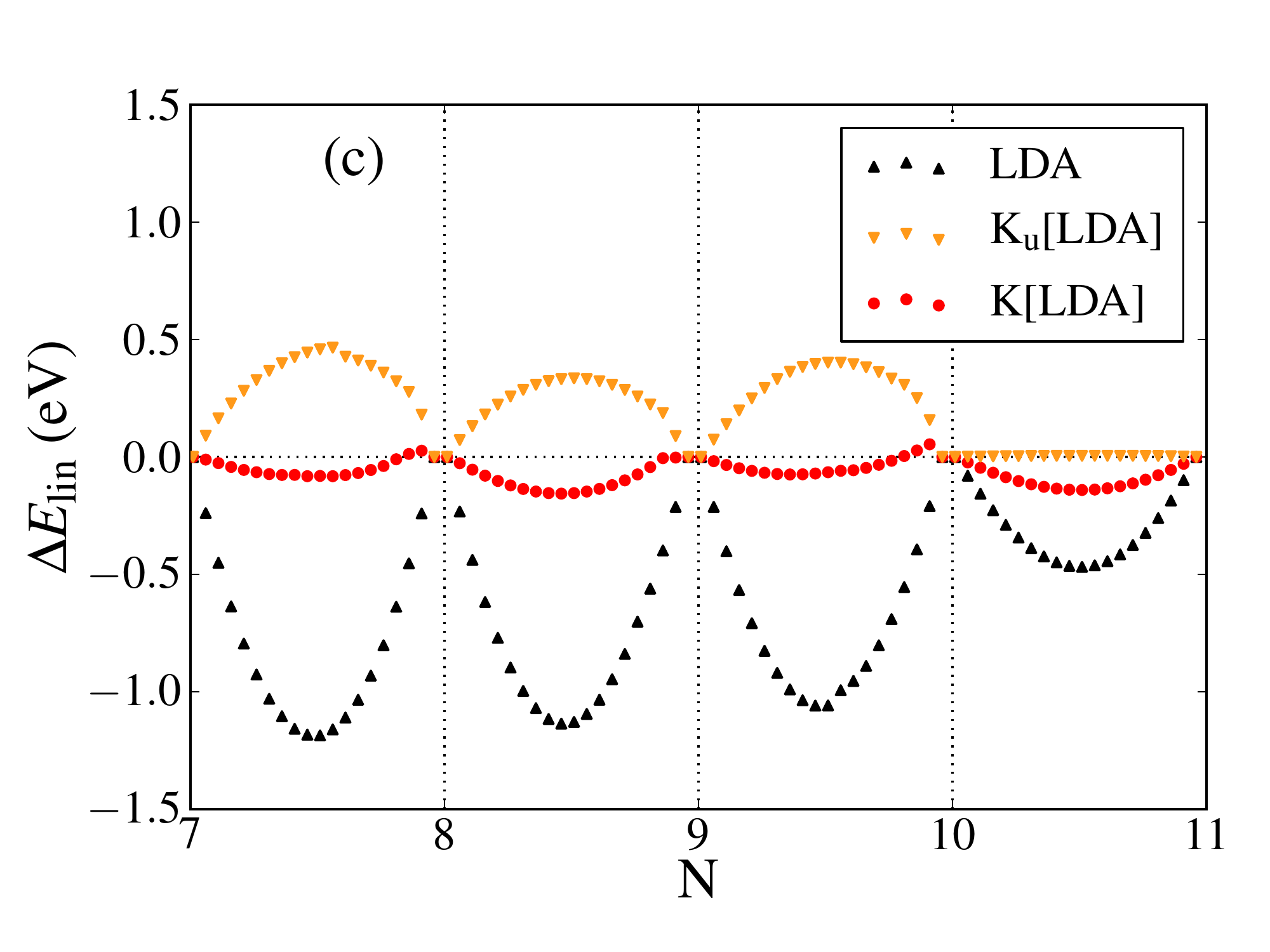}
\end{minipage}
\begin{minipage}{\linewidth}
\caption{(Color online) (a) HOMO eigenvalues as a function of electron number in methane using LDA, the unrelaxed K correction (K$_{\rm u}$), and the screened K correction. Eigenvalues have a finite slope as a function of occupation in LDA, which is reversed in K$_{\rm u}$, and almost disappears in K (dotted lines are a guide for the eye). The tiny kink in the red curve between 8 and 9 electrons, and the oscillation just below integer occupations 8 and 10 are the consequence of the localization of variational orbitals close to integer filling, an effect which was explained by Vydrov and Scuseria~\cite{Vydrov2007} in the case of PZ-SIC. (b) Integral of the quantity plotted on panel (a), showing how the piecewise linear dependence of energy vs. particle number is recovered within K (black straight lines are a guide for the eye and mark the piecewise linear behavior). The curves made of orange and black triangles have been shifted upwards for clarity. (c) Difference between each curve on panel (b) and the corresponding piecewise-linear behavior marked by the black straight lines. The LUMO of methane is unbound, and this is the reason for the apparently better performance of the unscreened K$_{\rm u}$ correction between 10 and 11 electrons.}\label{Fig:piecewise}
\end{minipage}
\end{figure}

It is a property of the exact energy of an isolated system at zero temperature to be piecewise linear as a function of particle number~\cite{perd+82prl}. As a consequence of Eqs.~\eqref{Eq:Janak} and~\eqref{Eq:Janak1}, the exact HOMO eigenvalue of such a system is piecewise constant as a function of the particle number $N_0$, with jumps at integer numbers. This means that, for the exact energy, the derivative in \eqn{Eq:Janak} equals the finite difference $I=E(N)-E(N-1)$, which defines the ionization energy~\cite{perd-levy83prl}. 
The property of piecewise linearity is usually not satisfied by $\epsilon_{\rm HO}$ in approximate DFT calculations, such as those based on local or semilocal functionals (see \fig{Fig:piecewise}). For such functionals, the value of $\epsilon_{\rm HO}$ is a (linear, to first order) function of $N$. As a result, $I^{\inbra{\rm LDA}}=E^{\inbra{\rm LDA}}(N)-E^{\inbra{\rm LDA}}(N-1)$ can be very different from $-\epsilon_{\rm HO}^{\inbra{\rm LDA}}(N)$. Indeed, deviations can be as large as several electronvolts. This flaw severely undermines the possibility of giving physical meaning to any LDA Kohn-Sham eigenvalue.
The purpose of Koopmans-compliant functionals is that of correcting this flaw, adding to the LDA energy a term which restores such piecewise linearity, by enforcing the equality
\begin{align}
E(f_{\rm HO}=1)-E(f_{\rm HO}=0) = -\epsilon_{\rm HO}(f_{\rm HO})\,,
\end{align}
for all values of $0<f_{\rm HO}<1$.
It is important to recall that the above piecewise linearity is never satisfied by the Perdew-Zunger orbital-density-dependent self-interaction correction~\cite{Perdew1981} (PZ-SIC) in many-particle systems. 
The reason is that PZ-SIC is designed to be exact for one-electron systems, where the self-interaction error is correctly defined as the interaction energy of a single electron. This definition is no longer valid for many-particle systems, where the interaction energy of a single particle has no longer any physical meaning. In order to see plots of the dependence of PZ-SIC energy derivative as a function of fractional occupation one can read for instance Refs.~[\onlinecite{Dabo2010,Vydrov2007}].

\subsection{Koopmans' correction}
Moving from \eqn{Eq:DFT_fract_def},
we choose to generalize \eqn{Eq:Janak1}, valid for the occupation of the HOMO, to the case of any occupation $f_i=s$ of an orbital $\phi_i$ (not necessarily an eigenstate),
\begin{align}\label{Eq:Janak_bis}
    \frac{\partial E^{\inbra{{\rm LDA}}}{(\{f\})}}{\partial f_i}\Big|_{f_i=s} = \eta_i(s)\,.
\end{align}
It is easy to show that, given the LDA Kohn-Sham Hamiltonian $\hat{H}_{\rm LDA}$, we have
\begin{align}\label{Eq:eta_def}
\eta_i(s)=\expect{\phi_i}{\hat{H}_{\rm LDA}(s)}\,.
\end{align}
If we had used a total derivative in \eqn{Eq:Janak_bis}, we would have had to take into account the relaxation of orbital $\phi_i$ in the final result, since \eqn{Eq:Janak1} proves that only the HOMO does not relax, to first order, upon changes in its occupation. 
%
We therefore maintain, in the following discussion, partial derivatives, and will show later in this section how to introduce the effects of orbital relaxation.
The Koopmans' correction is meant to impose the independence of $\eta_i(s)$ from $s$, in analogy with the independence of the eigenvalue of the HOMO from its occupation. One may remark that this condition is in principle more stringent than the constraint of piecewise linearity, which can be enforced by correcting the eigenvalue of the HOMO only, but the apparent excess of zeal is not left without reward.
Indeed, while the results of this paper are focused only on the prediction of ionization energies (from HOMO eigenvalues), other works show how the Koopmans' condition can result in successful predictions of electron affinities~\cite{Dabo2013} (from LUMO eigenvalues) and photoemission spectra~\cite{nguyen_photoemission} (from eigenvalues of lower-lying states).

We therefore define a new energy functional
\begin{equation}\label{Eq:unscreened-Koopmans}
E^{\inbra{{\rm KC}}}{(\{f\})} = E^{\inbra{{\rm LDA}}}{(\{f\})} + \sum_i \Pi_i(f_i)
\end{equation}
such that, for every $i$,
\begin{align}\label{Eq:Janak_bis2}
\frac{\partial E^{\inbra{{\rm KC}}}{(\{f\})}}{\partial f_i}\Big|_{f_i=s} = \bar{\eta}_i\,.
\end{align}
with $\bar{\eta}_i$ constant with respect to $s$.
In order to do this, we devise an orbital-dependent correction $\Pi_i(f_i)$ such that
\begin{align}\label{Eq:Janak_odd}
\frac{\partial \Pi_i(f_i)}{ \partial f_i}\Big|_{f_i=s} = -\eta_i(s)+\bar{\eta}_i\,,
\end{align}
i.e., its derivative removes the orbital energy dependence on occupations and replaces it with a constant $\bar{\eta}_i$ (see Sec.~\ref{sec:flavors} 
and \appendixx{App:KC}), whose expression in the case of the K functional follows Slater's intuition, and
is equal to $\eta_i(s=1/2)$. The value of $\eta_i(s)$ could be computed analytically, provided the manifold of wave functions $\{\phi_i(\br)\}$ is kept frozen to its value for $s=1$.
This leads to the so-called {\it frozen-orbital} (unscreened) {\it Koopmans' correction}, defined by the differential equation
\begin{align}\label{Eq:Janak_odd_0}
    \frac{\partial \Pi_i(f_i)}{\partial f_i}\Big|_{f_i=s,\{\phi\}} = -\eta^0_i(s)+\bar{\eta}_i\,,
\end{align}
where the superscript 0 on the right-hand side recalls the fact that the orbital manifold is defined at $s=1$ (it will be dropped in the following to simplify the notation).
%
%
Integrating \eqn{Eq:Janak_odd_0} from zero to $f_i$ at frozen orbitals, one obtains the frozen-orbital Koopmans' correction to the LDA energy
\begin{align}\label{Eq:KC_corr}
\Pi_i(f_i) &= -\int_0^{f_i} \eta_i(s) \de{s}+f_i\bar{\eta}_i \nonumber\\
&= E^{\rm LDA}[\rho-\rho_i] -E^{\rm LDA}[\rho]  + f_i \bar\eta_i\,,
\end{align}
[where $\rho_i(\br) = f_i|\phi_i(\br)|^2$], and the frozen-orbital Koopmans-compliant functional Eq.~\eqref{Eq:unscreened-Koopmans}.
The effect of orbital relaxation due to a change in fractional occupations is not taken into account by the above correction. This leads typically to a piecewise concave energy as a function of fractional number of particles, similarly to what happens in Hartree-Fock~\cite{perd+07pra}.
In order to achieve the desired piecewise linearity we need to take into account orbital relaxations, typically with a screening coefficient $\alpha$~\cite{Dabo2010}, yielding the {\it screened Koopmans-compliant functional}:
\begin{align}\label{Eq:KC_def}
E^{\inbra{{\rm KC}}}{(\{f\})} = E^{\inbra{{\rm LDA}}}{(\{f\})} +  \alpha \sum_i \Pi_i(f_i)\,,
\end{align}
which corresponds to the minimum of the functional introduced by~\eqn{Eq:Koopmans_introduction}, enforcing orthonormality constraints through a matrix of Lagrange multipliers $\Lambda_{ij}$.
The eigenvalues of this matrix provide a generalization of Kohn-Sham eigenvalues $\epsilon^{\inbra{\rm KC}}_i$ (see, e.g., Refs.~[\onlinecite{pede+84jcp,Vydrov2007,sten-spal08prb}] and Sec.~\ref{sec:variational_vs_canonical} for further
details). In \fig{Fig:piecewise} we show the effects of the unscreened and screened Koopmans' corrections on the energy and HOMO eigenvalue of methane as a function of fractional occupation.
The screening coefficient $\alpha$ is chosen such that the eigenvalue of the HOMO for the system with $N$ electrons is equal to the eigenvalue of the LUMO for the same system with $N-1$ electrons~\cite{Dabo2010, krai-kron13prl}, thus ensuring that the frontier orbital has an eigenenergy which does not change with fractional occupation, displaying the ``straight steps'' pattern shown in \fig{Fig:piecewise} (red circles). In this figure, the whole curve for K is shown for a single value of $\alpha$, computed for neutral methane.
It is important to stress that $\alpha$ is not a semiempirical parameter, but it is computed entirely from {\it ab initio} calculations (see~\sect{sec:screening} and in particular~\fig{Graph:alpha_calc} for details on the calculation of the screening). Another relevant remark is that, although $\alpha$ is computed from a condition on (relaxed) frontier orbitals (HOMO for $N$ particles and LUMO for $N-1$ particles), it proves to be an accurate estimate of orbital relaxation effects also for eigenstates with lower energy, thus enabling a reliable prediction of photoemission spectra~\cite{nguyen_photoemission} as well as ionization energies. 



\subsection{Different types of functionals}\label{sec:flavors}

\subsubsection{K and KI functionals}\label{Sec:K-KI}
There are two similar, meaningful choices for $\bar{\eta}_i$ in \eqn{Eq:Janak_odd_0}, which result in two different ``flavors'' of KC functionals.
For the first one, we can take Slater transition's state theory~\cite{Slater1974}:
\begin{align}
\bar{\eta}^{\inbra{\rm K}}_i\; =\; \eta_i(s=\fref)\,,
\end{align}
using $\fref=1/2$, which provides the K functional (labeled NK {in Ref.~[\onlinecite{Dabo2010}}]).
Alternatively, we can choose
\begin{align}\label{Eq:etabar_KI}
\bar{\eta}^{\inbra{\rm KI}}_i = \int_0^1 \eta_i(s) \de{s}\,,
\end{align}
which leads to a slightly different form of the correction that we refer to as KI (I standing for integral). 
It should be noted that (as rigorously proven in \appendixx{App:KC}) this correction restores piecewise linearity of the energy with respect to changes in particle number, but it does not change the LDA energy, nor the LDA ground state wave function (and consequently the one-body density-matrix) whenever the system has an integer number of particles.
The two corrections K and KI display scalar potential terms (i.e., contributions to the potential that
are constant over space) when taking the derivatives of the energy with respect to orbital densities $\rho_i(\br)$.
Since these terms do not depend on $\br$, they do not change the shape of the orbitals (see \appendixx{App:KC}), but only shift their energies.
We note that this breaks the relationship, existing in any KS-DFT calculation on finite systems, between the value of $\epsilon_{\rm HO}$ and the decay of the ground-state density away from the system, governed by the asymptotic equality\cite{perd+82prl, Almbladh1985}:
\begin{align}\label{Eq:longdistance_decay}
\log\sqbra{\rho(\br)}\underset{|\br| \to \infty}{\approx} -2r\sqrt{-2\epsilon_{\rm HO}} .
\end{align}
Broadly speaking, this means that in K and KI the eigenvalue of the HOMO is correct, but the total density preserves its incorrect decay away from the finite system.

\subsubsection{From K and KI to KPZ and KIPZ functionals}

The relation in Eq.~\eqref{Eq:longdistance_decay} between the density and the HOMO eigenvalue can be approximately restored by removing exactly the scalar potential arising from the Hartree part of the KC correction, and approximately the scalar potential arising from the exchange-correlation part of the KC correction. This can be achieved by combining the PZ-SIC energy with the K and KI orbital-density-dependent corrections, therefore obtaining what we will label KPZ and KIPZ functionals, respectively. As better discussed in~\appendixx{App:KL_KIL}, these functionals have the important property of being exact for one-electron systems, while at the same time being able to preserve the piecewise linearity of the energy in many-electron systems, and thus are exactly free from the one-body SIE, and approximately free from the many-body SIE~\cite{cohe+08sci}.

We now proceed to explain how to obtain the KPZ or KIPZ functionals by linking these seamlessly to K or KI. For this purpose we can define, merely as mathematical tools, the K$\mhs{L}$ and KI$\mhs{L}$ functional corrections; for K$\mhs{L}$ we have
\begin{align}\label{Eq:Kl_def}
\Pi^{\inbra{\rm \gamma K\mhs{L}}}_i(f_i) = \gamma \Pi^{\inbra{\rm PZ}}_i-\int_0^{f_i} \eta^{\inbra{{\rm \gamma PZ}}}_i(s) \de{s} + f_i\bar{\eta}^\inbra{\rm \gamma K\mhs{L}}_i\,,
\end{align}
where
\begin{eqnarray}
    \label{Eq:PZ}
    \Pi^{\inbra{\rm PZ}}_i\sqbra{f_i} &=& {-\EHxc\sqbra{\rho_i} }\,,
    \\[7pt]
    \label{Eq:Janak_bis0}
    \eta^{\inbra{{\rm \gamma PZ}}}_i(s) &=& \frac{\partial \cubra{E^{\inbra{{\rm LDA}}} + 
          \gamma \sum_j \Pi^{\inbra{\rm PZ}}_j}}{\partial f_i}\Big|_{f_i=s}\,,
    \\[7pt]
    \bar{\eta}^\inbra{{\rm \gamma K}\mhs{L}}_i &=& \eta^{\inbra{{\rm \gamma PZ}}}_i(\fref)\,,
\end{eqnarray}
and where $\rho_i(\br)=f_i |\phi_i(\br)|^2$.
Similarly to \eqn{Eq:etabar_KI}, the KI$\mhs{L}$ functional can be defined using the same equation for K$\mhs{L}$, \eqn{Eq:Kl_def}, modified with a different definition for $\bar{\eta}$:
\begin{align}
\bar{\eta}^\inbra{{\rm \gamma KI}\mhs{L}}_i = \int_0^1\eta^{\inbra{{\rm \gamma PZ}}}_i(s)\de{s}\,.
\end{align}

The parameter $\gamma$ tunes the weight of the Perdew-Zunger correction [\eqn{Eq:PZ}] relatively to the KC correction.
When $\gamma\to 0$ one recovers K and KI from K$\mhs{L}$ and KI$\mhs{L}$, respectively; on the other hand, for $\gamma\to 1$ one obtains KPZ and KIPZ. In this work we show results only for extremal values of $\gamma=0$ and $\gamma=1$.
An important remark is that in the case of the KI functional, our results will be defined as the limit for $\gamma\to 0$ of KI$\mhs{L}$, and not as KI$\mhs{L}$ for $\gamma$ identically equal to zero. 
We need to adopt the subtlety of this limiting procedure in order to remove an ambiguity in the definition of orbital densities $\rho_i(\br)$ for KI. Indeed, since the energy of KI matches exactly the value of the unitary invariant LDA functional for integer number of particles, there is no way to select a unique set of orbitals $\phi_i$ by energy minimization (they can still be mixed by a unitary rotation at no energy cost), unless an infinitesimal value of $\gamma$ is introduced.

The limiting procedure described above has a negligible effect
on the calculation of KI electronic eigenvalues in small molecules, in which orbitals remain strictly localized in space, but becomes crucial when correcting KI band gaps of extended molecules and crystal systems, in which this procedure enforces the localization of orbital densities.
\subsubsection{The K0 non variational functional}
Since the scalar potential terms resulting from the K correction were early identified as by-products of the variational minimization, a non variational flavor, called K0, without these contributions had been introduced~\cite{Dabo2009}.
As better explained in \appendixx{App:K0}, and by the results of Dabo \etal~\cite{Dabo2010} (where it is termed NK$_0$), the K0 correction can be seen as a non variational form derived from the K energy, where the potential does not include the derivative of the energy with respect to a change in Slater's transition-state wave function, nor the derivative of each orbital-density-dependent energy correction $\Pi_i$ with respect to changes in orbital densities $\rho_j(\br)$ with $j\neq i$ (cross derivatives). 
As one of its main features, the K0 potentials ${V}^i(\br)$ do not contain scalar terms, and correspond to orbital-dependent Hamiltonians $\hat{H}^i$ each of which is identical to the LDA (or PBE) Hamiltonian for a system of $N-1/2$ electrons, half an electron having been removed from the wave function $\phi_i(\br)$. In practical calculations, K0 non variational results are obtained from a damped-dynamics minimization of the electronic wave functions driven by the K0 potentials, and subject to orthonormality constraints.
\subsubsection{All variational functionals at a glance}
Here we summarize all the expressions for the variational Koopmans-compliant functionals introduced in the previous paragraphs. These are the K functional, for which $E^{\rm K} = E^{\rm LDA} + \sum_i \Pi_i^{\rm K}$, KI, with $E^{\rm KI} = E^{\rm LDA} + \sum_i \Pi_i^{\rm KI}$, KPZ, with $E^{\rm KPZ} = E^{\rm LDA} + \sum_i \Pi_i^{\rm KPZ}$, and KIPZ, with $E^{\rm KIPZ} = E^{\rm LDA} + \sum_i \Pi_i^{\rm KIPZ}$.
In the following we recall, with a notation involving Hamiltonian matrix elements, and totally equivalent to the one provided in previous sections, the values of all corrections $\Pi_i$:
\begin{widetext}
\begin{align}\label{Eq:K_sumup}
\Pi^{\rm K}_i &= -\int_0^{f_i} \expect{\phi_i}{\hat{H}_{\rm LDA}(s)} \de{s}+f_i\expect{\phi_i}{\hat{H}_{\rm LDA}(1/2)} \,,\\
\Pi^{\rm KI}_i &= -\int_0^{f_i} \expect{\phi_i}{\hat{H}_{\rm LDA}(s)} \de{s}+f_i\int_0^{1}\expect{\phi_i}{\hat{H}_{\rm LDA}(s)}\de{s} \,,\\
\Pi^{\rm KPZ}_i &= \Pi^{\rm PZ}_i(f_i)-\int_0^{f_i} \expect{\phi_i}{\hat{H}^i_{\rm PZ}(s)} \de{s}+f_i\expect{\phi_i}{\hat{H}^i_{\rm PZ}(1/2)}\,,\\
\Pi^{\rm KIPZ}_i &= \Pi^{\rm PZ}_i(f_i)-\int_0^{f_i} \expect{\phi_i}{\hat{H}^i_{\rm PZ}(s)} \de{s}+f_i\int_0^{1}\expect{\phi_i}{\hat{H}^i_{\rm PZ}(s)} \de{s}\,,
\end{align}
\end{widetext}
where 
\begin{align}
&\expect{\phi_i}{\hat{H}^i_{\rm PZ}} = \expect{\phi_i}{\hat{H}_{\rm LDA}}\nonumber\\
&-\sum_i \int \phi^\ast_i(\br)\vHxc(\br; [\rho_i])\phi_i(\br) \de{\br}\,,
\end{align}
and where $\vHxc(\br; [\rho_i])$ is the PZ-SIC potential for the $i_{\rm th}$ orbital, i.e, the Hartree plus the exchange-correlation potential for the orbital density $\rho_i(\br)$.

\section{Results}\label{Sec:results}

\begin{figure}
\includegraphics[width=0.45\textwidth]{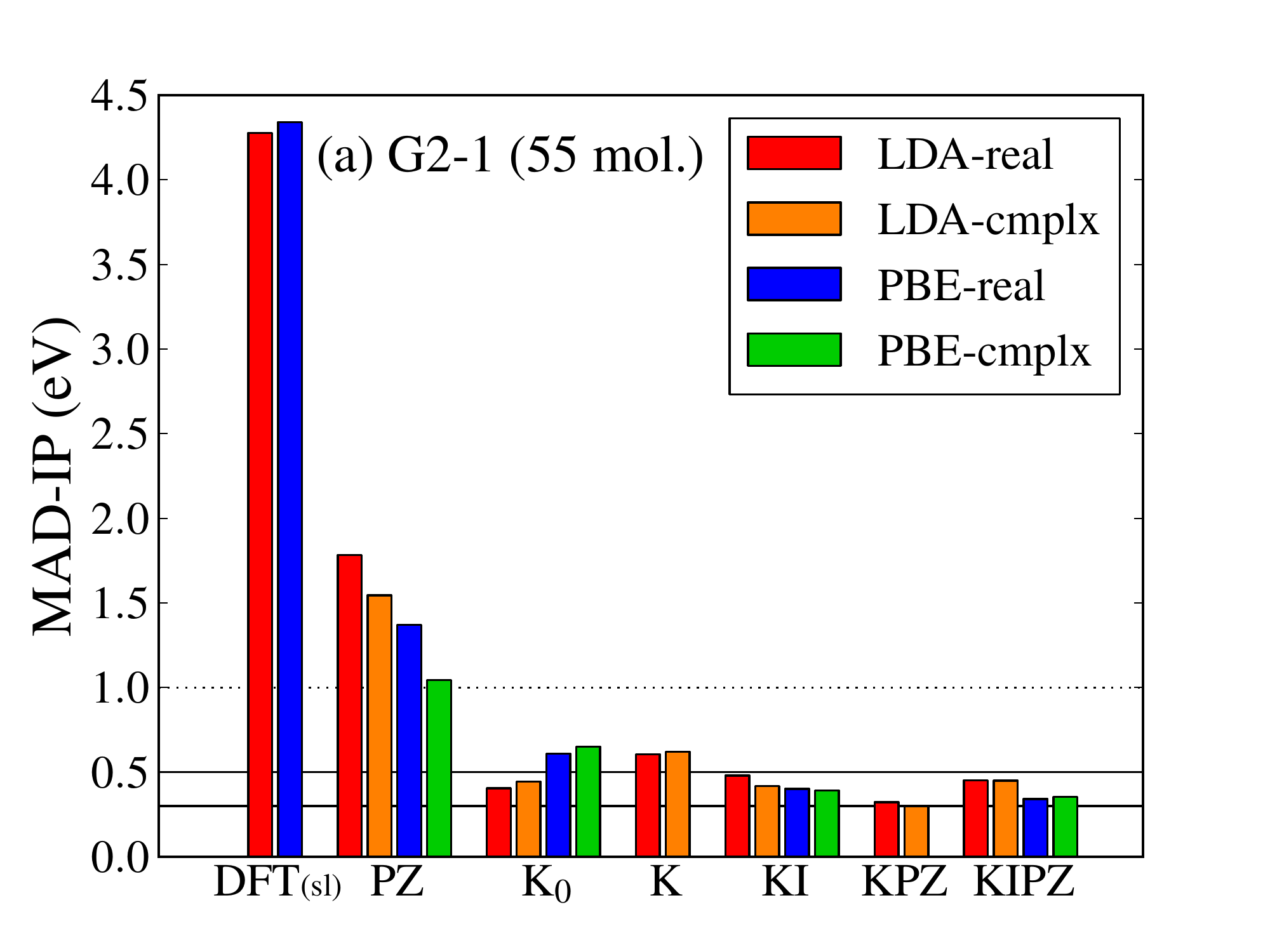}
\includegraphics[width=0.45\textwidth]{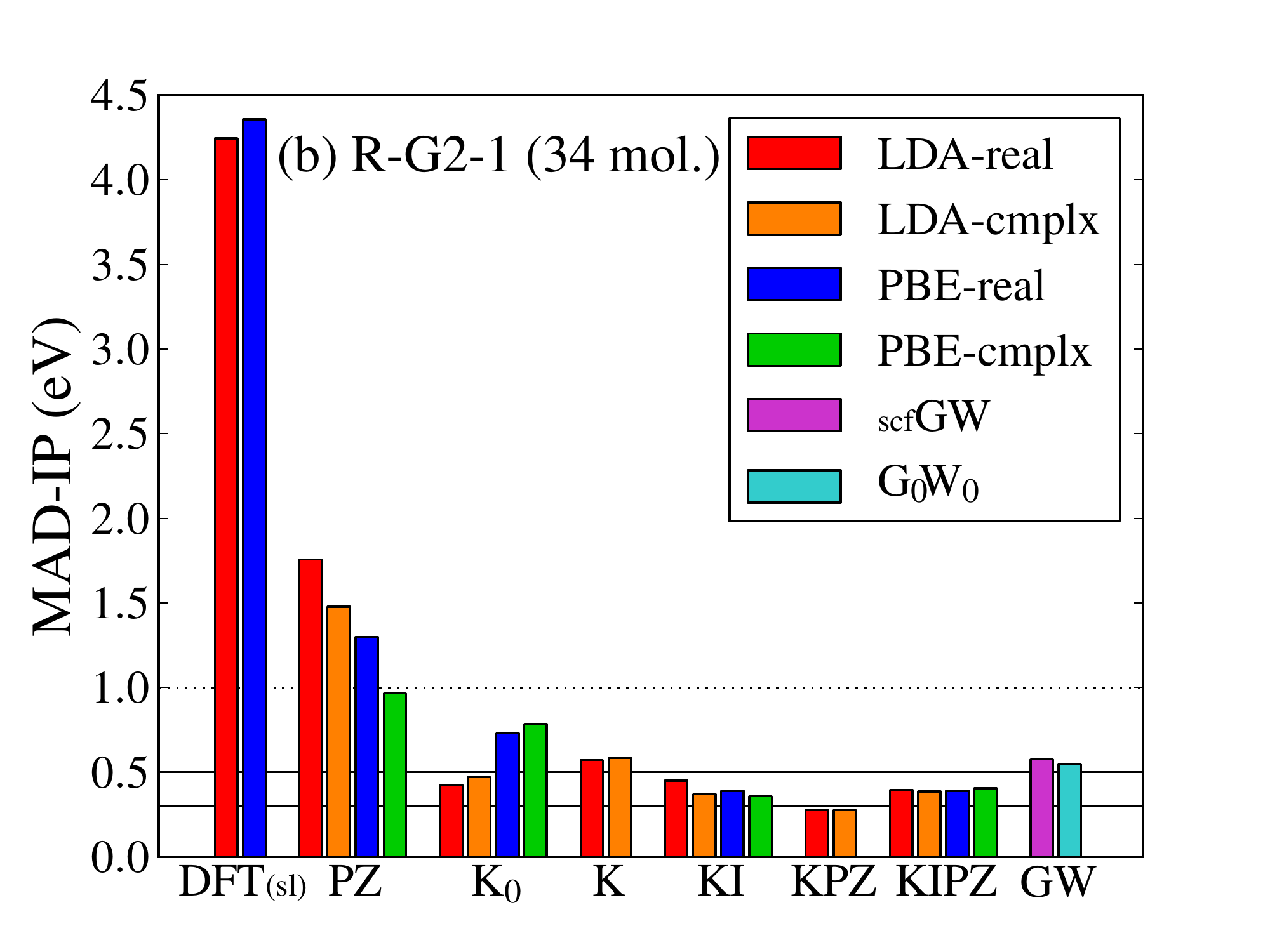}
\caption{(Color online)
   (a) Upper panel: mean absolute deviation of DFT, PZ, and KC ionization energies from experimental values, 
   averaged over the 55 molecules of the full G2-1 test set, computed for all KC functional flavors including the K0 non variational approximation, 
   for real- and complex-valued minimizing orbitals and for LDA and PBE exchange-correlation functionals. 
   (b) Lower panel: comparison of our results with the self-consistent GW ({\scriptsize scf}GW) and G$_0$W$_0$ results of Ref.~[\onlinecite{Rostgaard2010}], 
   performed over their restricted R-G2-1 set of 34 molecules.
   In both panels, the lowest horizontal black solid line marks the value of 0.3~eV. An enlarged view of the results for ODD functionals and GW is available in the Supplemental Material~\cite{Supplemental_material} (SM Fig.~1).}\label{Fig:bar_mr}
\end{figure}

\begin{figure}
\includegraphics[width=0.40\textwidth]{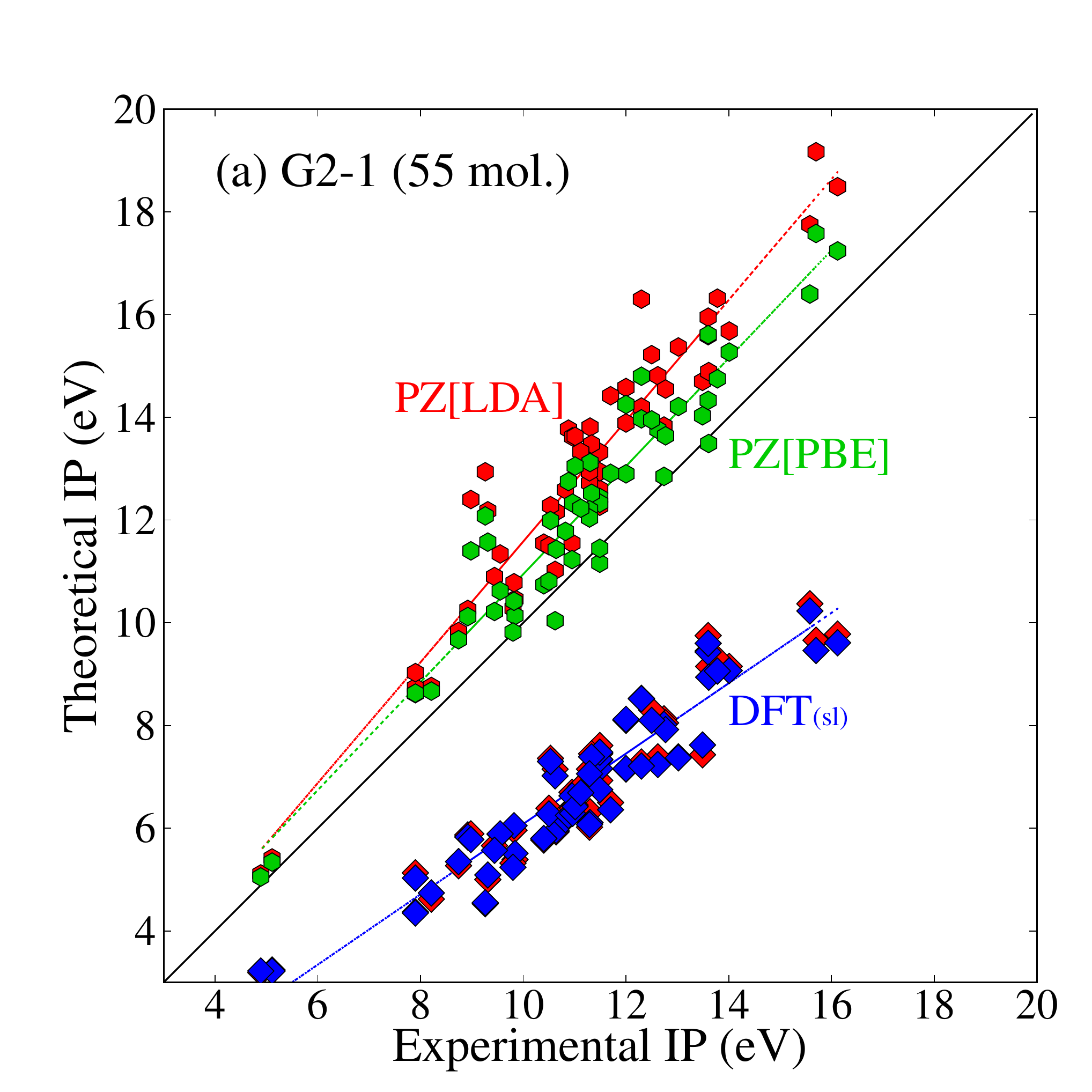}
\includegraphics[width=0.40\textwidth]{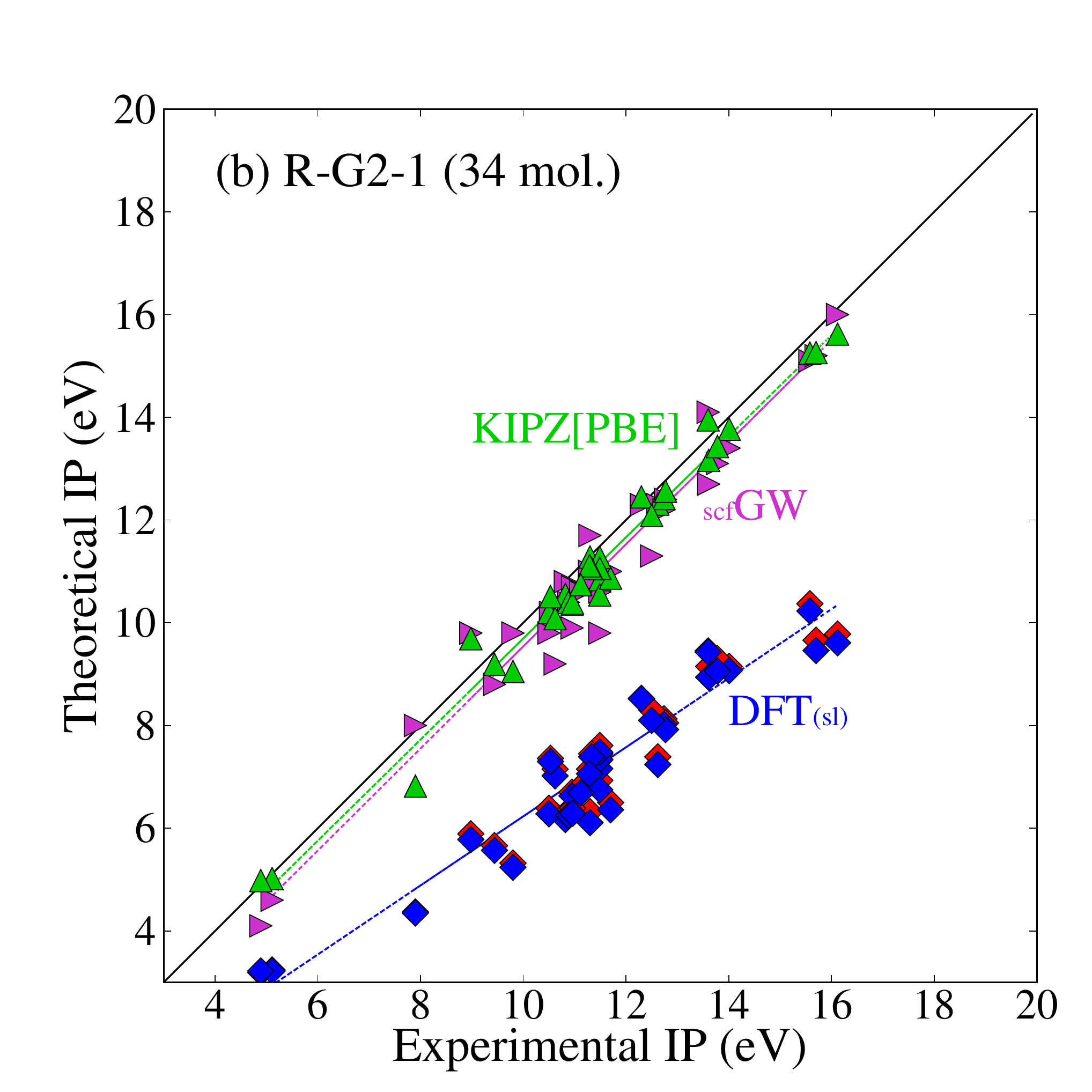}
\caption{(Color online) Theoretical versus experimental ionization energies for the G2-1 set.
  (a) Upper panel: 
  LDA (red diamonds) and PBE (blue diamonds) results together with the least-squares fit for PBE data (blue line). 
  Results from PZ-SIC functional on top of LDA using real wave functions 
  (red hexagons and red line as a least-squares fit of the data), and on top of PBE minimized using complex orbitals 
  (blue hexagons and blue line) are also shown. 
  (b) Lower panel: LDA and PBE data for the restricted R-G2-1 set, together with results from the 
   KIPZ functional on top of PBE minimized using complex orbitals (green triangles and green line) 
  and the reference self-consistent GW data (violet triangles and line) from Ref.~[\onlinecite{Rostgaard2010}].}
  \label{Fig:line_pz_nkipz_g0w0_dft}
\end{figure} 

As mentioned in \sect{Sec:KCdef}, the eigenvalue of the highest-occupied molecular orbital in DFT calculations with local and semilocal functionals such as LDA and PBE is unable to provide a reliable estimate for the ionization energy of a system. Given that these functionals provide good total energies at integer occupations, the main cause of this flaw is the lack of piecewise linearity in the energy versus fractional occupation. 

In this section we will therefore show how the KC orbital-density-dependent functionals are able to restore the reliability of $-\epsilon_{\rm HO}$ as an estimate of the ionization energy. We will consider KC corrections applied both on top of LDA and (for all flavors except K and KPZ~\footnote{It is possible in principle to compute K and KPZ results on top of PBE, provided the PBE exchange-correlation kernel can be computed. The calculation of the PBE kernel is delicate, and we found that in our code it was a source of numerical instability.}) of PBE KS data. The minimization of all functionals have been performed on either real-valued and complex-valued wave functions, and the results of the minimizations on the two different sets will be compared. After the calculation of ionization energies, we will consider also the effect of orbital-density-dependent corrections (PZ or KC) on the geometries of most molecules in the G2-1 set, and we will investigate the performance of PZ and KC functionals in the calculation of atomization energies.

All experimental results are taken from reference data available on the NIST website~\cite{G2_web_data}, or equivalently from 
Refs.~[\onlinecite{curtiss_reference_g2},\onlinecite{staroverov_g2_reference}]. 
When comparing with the self-consistent GW and non-self-consistent 
G$_0$W$_0$ results of Rostgaard \etal~[\onlinecite{Rostgaard2010}], we consider their same 
restricted version of the G2-1 set (labeled as R-G2-1 in the figures) contaning 34 molecules instead of the 
original 55 ones. 

\subsection{Details on the calculations}\label{Sec:computational}

All calculations are performed with a modified version of the Car-Parrinello code in the Quantum-ESPRESSO distribution.
The implementation is based on a plane-wave basis set using LDA~\cite{Perdew1981} and PBE~\cite{Perdew1996} norm-conserving pseudopotentials. A discussion about the error introduced in Hartree-Fock and ODD calculations with the use of LDA and PBE pseudopotentials can be found in Ref.~[\onlinecite{Dabo2010}], in which it is also possible to see (in section IIIA) the results of atomic all-electron calculations performed with the K functional. The error caused by pseudopotentials in the calculation of ionization energies is estimated to be around 0.1--0.2~eV.
For the calculation of the $\alpha$ screening coefficients, we have followed the scheme described in Ref.~[\onlinecite{Dabo2010}], which is recalled also in~\sect{sec:screening} of this paper. 

We use tabulated geometries for all G2-1 set~\cite{G2_web_data} molecules, and we set a kinetic energy cutoff of $60$~Ry for the wave functions and of $240$~Ry for the charge density. Each molecule is placed inside an orthorhombic cell having linear dimensions such that at least 18~Bohr of vacuum separates molecular replicas. This separation is sufficient to converge the total energy and the electronic eigenvalues since the Coulomb interaction between periodic images is suppressed by means of reciprocal space counter-charge corrections~\cite{Li2011}.
Once screening coefficients are computed, we use them to evaluate ionization and atomization energies, increasing the cutoff to 100~Ry and the vacuum size to 20~Bohr. 
The results reported for molecular geometries (bond lengths and angles) are obtained by starting from 
the tabulated G2-1 geometries (and the initially computed screening coefficients) and by performing 
a Car-Parrinello damped-dynamics structural optimization for each molecule and each functional. 
The cutoff and vacuum size used for geometry optimizations are 60~Ry and 20~Bohr, respectively.

In general, the spin configurations adopted are $s=0$($s=1/2$) for even- (odd-) electron molecules, while Hund's 
rules have been used for choosing the spin imbalance of atoms, required for atomization energy calculations. 
There are a few exceptions among the molecules, for which the lowest-energy spin configuration is not the one obtained with the above recipe: O$_2$ and S$_2$ have $s=1$, as well as triplet CH$_2$ and SiH$_2$, while the singly ionized molecules (necessary to compute the screening coefficient) NH$_2$ and OH have also $s=1$. 

In all plots that follow we will use the notation SiH$_2$\_s1A1d and SiH$_2$\_s3B1d, and the analogous notation for CH$_2$, to refer to the singlet and triplet spin configurations, respectively.
In the Supplemental Material~\cite{Supplemental_material} we display more figures presenting our results.
\subsection{Ionization energies}
In Figs.~\ref{Fig:bar_mr}(a) and~\ref{Fig:bar_mr}(b),
we show a comparison between the MAD (mean absolute deviation) of experimental ionization energies and theoretical 
data as obtained with all DFT and orbital-density-dependent functionals for the whole 55-molecule G2-1 set (upper panel) 
and for a subset of it (34-molecule, lower panel), which was the object of a self-consistent GW (scf-GW) study by Rostgaard \etal~\cite{Rostgaard2010}. Tabulated data for all ionization energies used in this figure can be found in the Supplemental Material~\cite{Supplemental_material} (SM Tables I to IV).
Two main remarks can be made by looking at these results.

The PZ (we will use this shorter acronym in figures to indicate PZ-SIC) error always exceeds the eV, while the KC errors are, in the worst case, half as large. 
Using a complex wave-function manifold and a gradient-corrected exchange-correlation functional is crucial to obtain a good PZ estimate of the ionization energy. 
Overall, one notices the great improvement of KC schemes over PZ results, such that a precision comparable or larger than that of G$_0$W$_0$ and scfGW calculations is achieved. 
The accuracy of KC functionals is emphasized also in \fig{Fig:line_pz_nkipz_g0w0_dft} (bottom panel), where the results for each single molecule of the 34-molecule R-G2-1 subset are shown for the KIPZ functional and G$_0$W$_0$. Least-squares fits act as a guide to the eye, and show the large discrepancy beween LDA and PBE results and experiment, which is also evident in the wrong slope of the fit for the PBE results. In the Supplemental Material~\cite{Supplemental_material} (SM Fig.~2), we present KI and KIPZ results analogous to those on \fig{Fig:line_pz_nkipz_g0w0_dft}(b) for the 55 molecules of the full G2-1 set, as well as (SM Fig.~3) the distribution of deviations from experiment of ionization energies computed with KI and KIPZ and compared to the ones found with G$_0$W$_0$ and scf-GW, which further assess the reliability of KC schemes.

As a second remark, we point out the strong difference between PZ results obtained when correcting LDA or PBE functionals, and when minimizing the energy on the set of real or complex wave functions. The better accuracy of the PZ results for ionization energies when computed by minimization on the space of complex wave functions and on top of the PBE functional was also discussed for atoms by Kl\"upfel \etal~\cite{Klupfel2011}, and later investigated for a group of five molecules~\cite{Hofmann2012}.
\subsection{Geometry optimization}

In order to assess the effectiveness of Koopmans-compliant schemes in predicting molecular geometries, we optimize the structure of all molecules of the G2-1 set, keeping the screening coefficients fixed to the values found in the previous section. We optimize the geometries within all functional schemes except for the K0 scheme, which is non variational. 

\begin{figure}
\includegraphics[width=.45\textwidth]{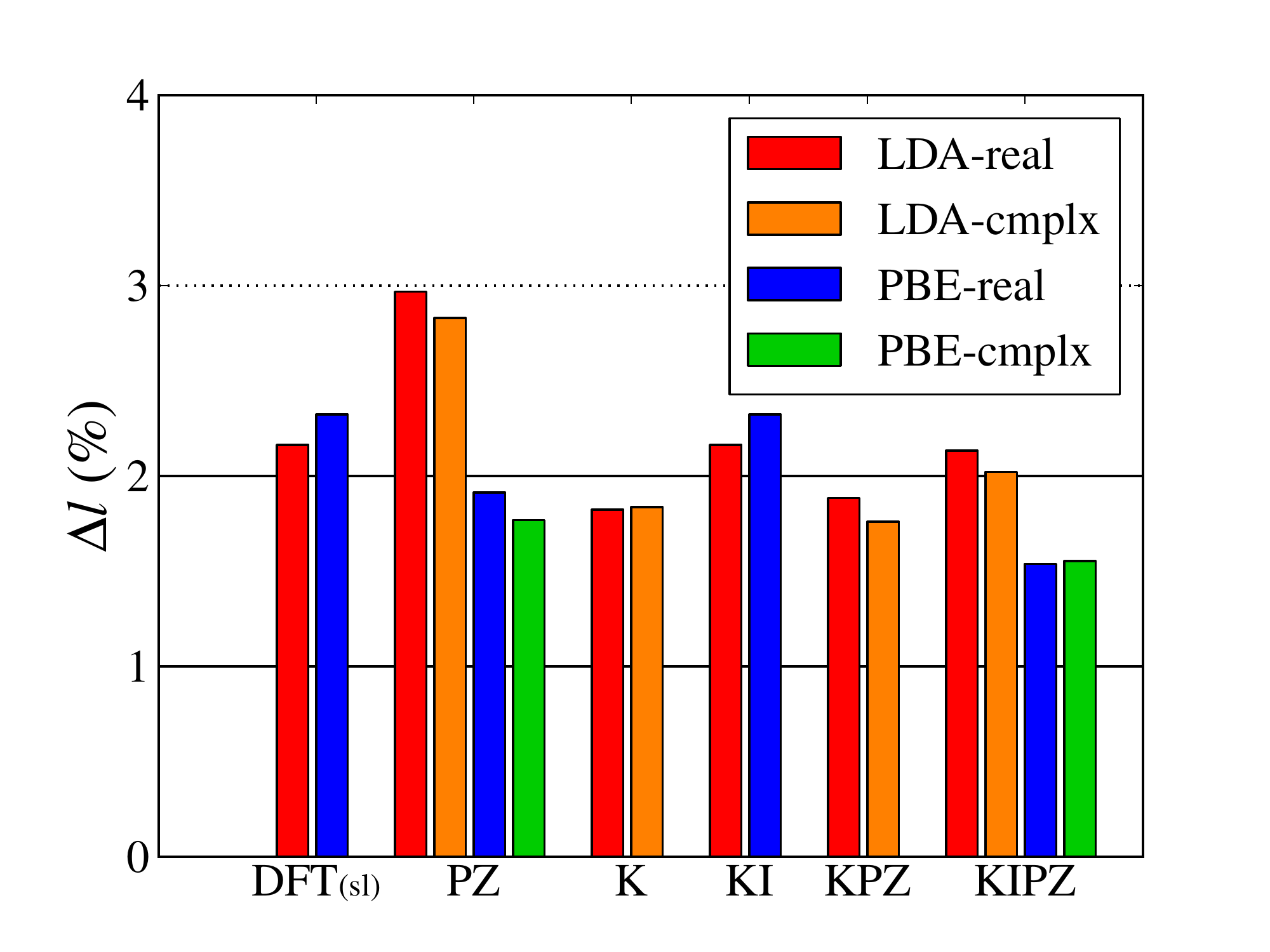}
\caption{(Color online). Absolute value of the bond-length relative error with respect to experiments averaged over all dimers in the G2-1 set, including 
     H$_2$ (P$_2$, SO, SiO, CS, NH, S$_2$, NO, Si$_2$, N2, O2, H$_2$, NaCl, CN, OH, ClO, ClF, F$_2$, HF, CH, 
     BeH, LiH, LiF, HCl, Li$_2$, Na$_2$, Cl$_2$, CO), and computed for all variational functionals 
     described in this paper. 
     }\label{Fig:alats_dimers_percent}
\end{figure}

\begin{figure}
\includegraphics[width=0.45\textwidth]{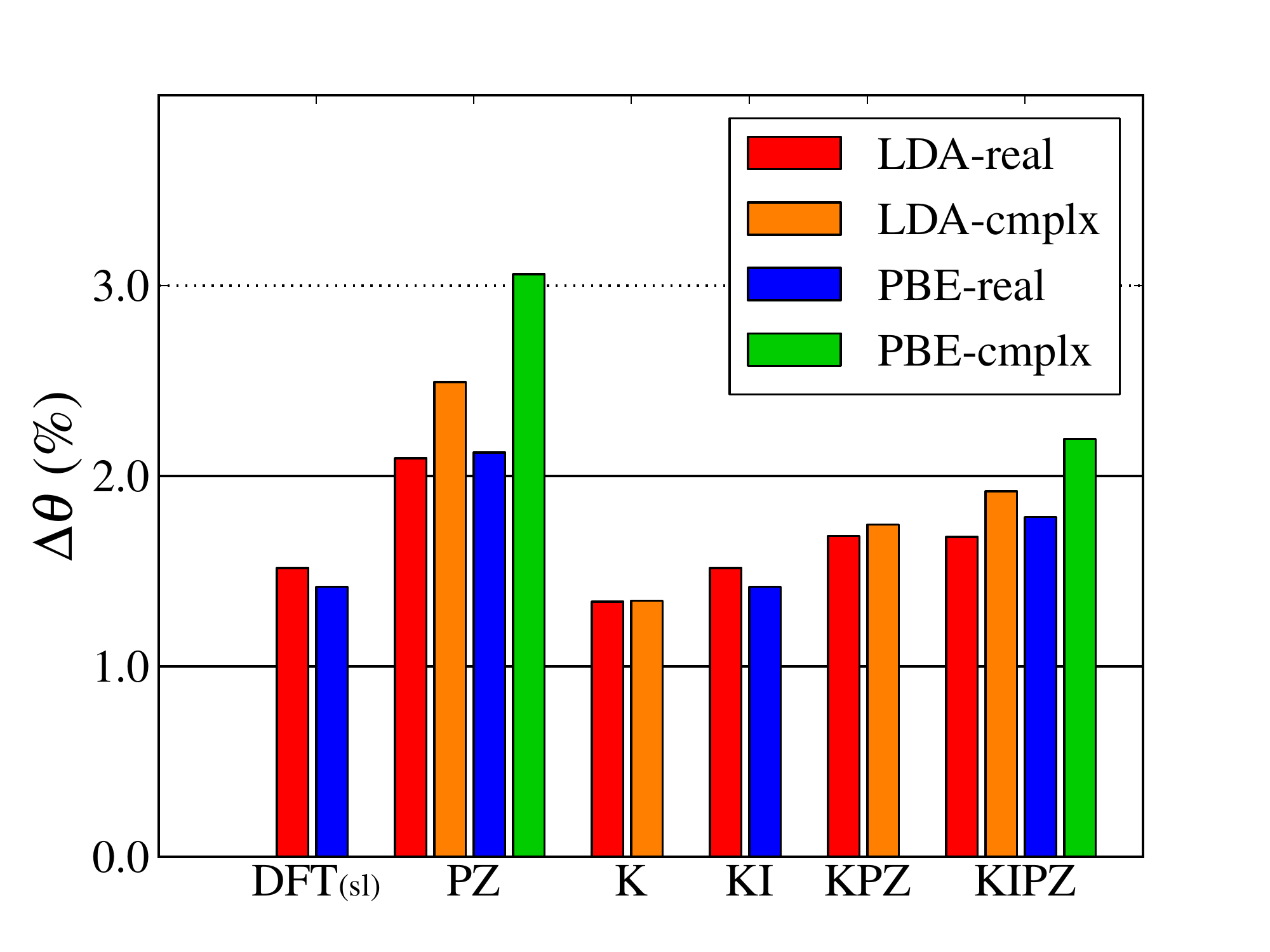}
\caption{(Color online) Plot of $\Delta\theta^{(\rm k)}$ (quantifying the angle relative error) as defined in~\eqn{Eq:Angle_error}, 
       averaged over all trimers and tetramers in the G2-1 set (NH$_2$, NH$_3$, H$_2$O, CH$_2$, SO$_2$, 
       PH$_2$, PH$_3$, SH$_2$, SiH$_2$, SiH$_2$, H$_2$CO, H$_2$O$_2$, HCO, HOCl). 
       For each molecule, only its smallest angle has been used in the average. 
       }\label{Fig:angles_trimers_tetramers_percent}
\end{figure}

Our results are summarized in \fig{Fig:alats_dimers_percent} for the bond length of the dimers in the set, and in \fig{Fig:angles_trimers_tetramers_percent} for the bond angle of trimers and tetramers. In the first figure, we plot the average of the absolute values of percentage deviation from experiment, i.e., the quantity
\begin{align}\label{Eq:bond_error}
\Delta l^{(\rm k)}= \frac{1}{N_{\rm mol}} \sum_{j=1}^{N_{\rm mol}} \Bigg|\frac{l^{(\rm k)}_{j,\rm calc} -l_{j,\rm expt}}{l_{j,\rm expt}}\Bigg|\,,
\end{align}
with $l_{j,\rm calc(expt)}$ being the calculated (experimental) bond length of dimer $j$, and $\rm k$ being a label for the functional used.
According to our results, the KIPZ scheme on top of the PBE functional appears to be the best candidate 
for the prediction of bond lengths, with an average deviation about 1.5\%, as well as for ionization energies,
where $\Delta$ IP is about 0.3--0.4 eV.

In Figs.~4 and~5 of the Supplemental Material~\cite{Supplemental_material} and the related discussion in the captions we provide some further insight concerning the performance of KC functionals in predicting bond lengths by showing the distribution of deviations from experiments, together with all theoretical predictions for the H$_2$ molecule.
For angles, we plot instead the quantity
\begin{align}\label{Eq:Angle_error}
\Delta\theta^{(\rm k)}=\frac{1}{N_{\rm mol}} \sum_{j=1}^{N_{\rm mol}} \Bigg|\frac{\theta^{(\rm k)}_{j,\rm calc}-\theta_{j,\rm expt}}{\theta_{j,\rm expt}}\Bigg|\,.
\end{align}
which enables us to conclude that all ODD flavors except K show a larger average deviation from experiment than LDA and PBE, but also that KC functionals do slightly improve over the PZ-SIC functional also when predicting molecular angles. Figure~6 of the Supplemental Material~\cite{Supplemental_material} shows all functional predictions for the angle between the two OH bonds in the water molecule.
\subsection{Atomization energies}
\begin{figure}
\includegraphics[width=0.45\textwidth]{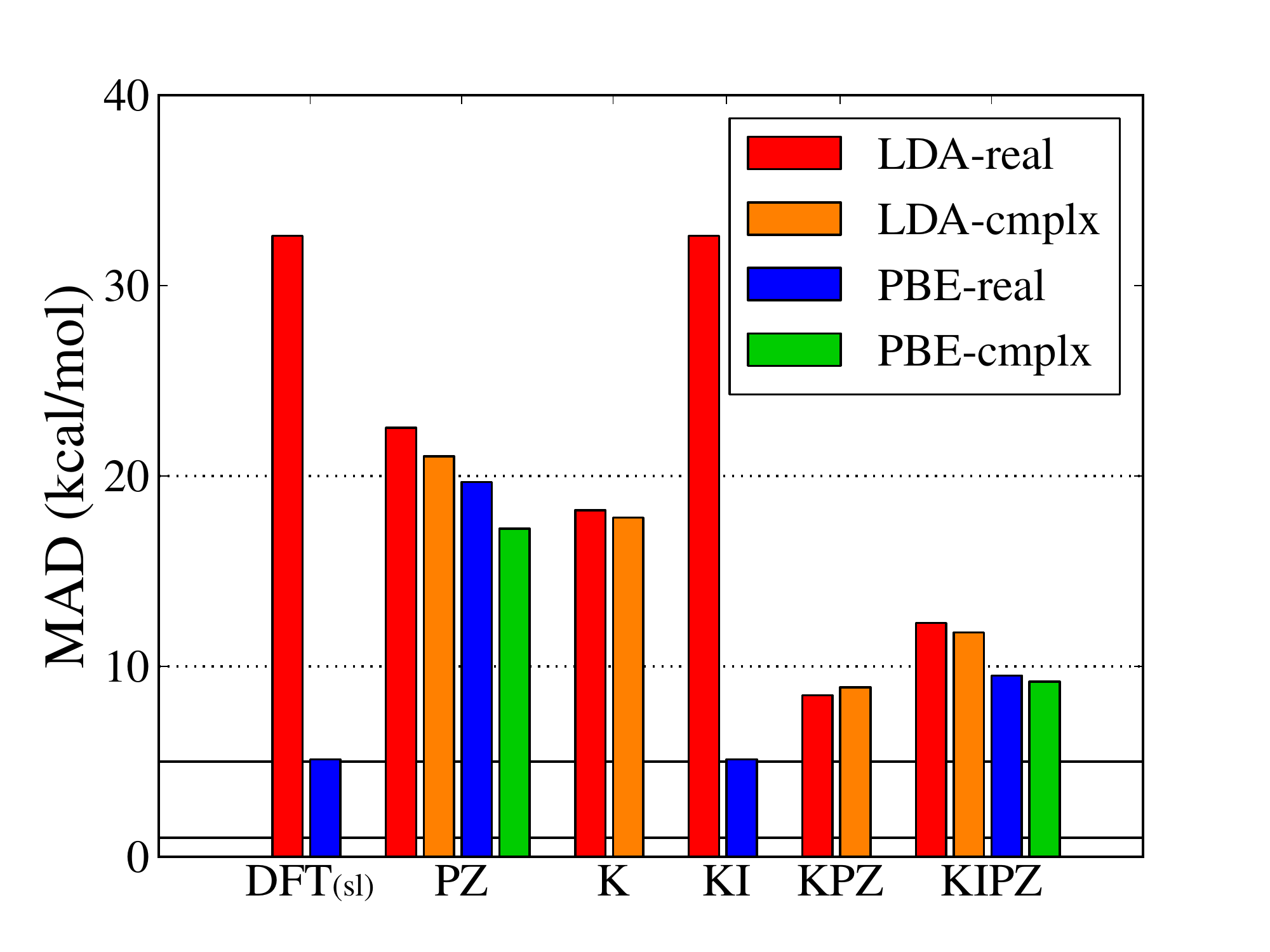}
\caption{(Color online) Mean absolute deviation from experiment of atomization energies of molecules in the G2-1 set except triplet SiH$_2$ (SiH$_2$\_s3B1d) and singlet CH$_2$ (CH$_2$\_s1A1d), and including H$_2$. Results are for all functionals discussed in this paper.
}\label{Fig:atomization_errors}
\end{figure}

Aside from the calculation of optimized geometries, another way of understanding the performance of all the ODD schemes described in this paper concerning total energies is the calculation of atomization energies. For dimers, this energy coincides with the binding energy. It is a well-known issue of LDA the fact that it severely overestimates binding energies, mainly because of its poor description of localized states of atoms. This overbinding of LDA can be overcome satisfactorily through the addition of gradient corrections, and indeed the PBE functional provides much more accurate estimates of binding and atomization energies of molecules. Figure~\ref{Fig:atomization_errors} 
shows that unfortunately neither the PZ functional, nor KC functionals are able to predict atomization energies better than PBE. The KI functional on top of PBE is by definition as good, while among the others we can say that the KPZ flavor, together with the KIPZ flavor on top of PBE and minimized with complex wave functions provide the second most accurate results.
In the Supplemental Material~\cite{Supplemental_material} (SM Fig.~7) we provide a plot of the distribution of deviations of theoretical from experimental atomization energies, which further supports these statements, showing in particular the improvement that the KIPZ functional brings about with respect to PZ-SIC, which was already shown~\cite{Klupfel2012} to have a sizable underbinding bias in molecules.

\section{Technical aspects}\label{Sec:technical}

We devote this section to further discuss some technical issues concerning the
definition and the implementation of ODD functionals. 
Namely, in Sec.~\ref{sec:variational_vs_canonical} we clarify the nature of 
the minimizing (variational) orbitals as compared to the so-called canonical ones (those diagonalizing the matrix of
Lagrange multipliers).
Then in Sec.~\ref{sec:OOD_per_atom} we address the dependence
of the ODD total energy correction on the degree of localization of the variational orbitals. We also discuss
the role of complex wave functions in the minimization of ODD functionals [Sec.~\ref{sec:cmplx_vs_real}]. Finally
in Sec.~\ref{sec:screening} we review the properties of the computed screening coefficients $\alpha$.

\subsection{Variational vs canonical orbitals}
\label{sec:variational_vs_canonical}

As mentioned in Sec.~\ref{Sec:KCdef}, at variance with density functionals, orbital-density-dependent functionals
can break the invariance of the total energy against unitary rotations of the (occupied) orbitals.
This is the case for PZ-SIC and, to a lesser extent, all KC flavors except for KI with orbitals at full occupations, for which, as pointed out in \sect{sec:flavors}, the energy remains identical to the one of the original (LDA or PBE) functional.
Let us consider the transformation
\begin{equation}
   \label{eq:mixingU}
   | \phi_j' \rangle = \sum_i U_{ij} | \phi_i \rangle .
\end{equation}
At the energy minimum, keeping the orbitals fixed but allowing for a unitary rotation among them, 
requires the condition
\begin{equation}
   \label{eq:derivU}
   \frac{\partial E^{\text{ODD}}}{\partial U_{ij}} = 0,
\end{equation}
which is non-trivially fulfilled (at variance with KS-DFT). In fact, the above equation
actually defines the specific unitary rotation leading to the energy minimum.
When considering the manifold of the occupied orbitals, Eq.~(\ref{eq:derivU}) can be cast
into~\cite{sten-spal08prb} 
\begin{equation}
   \label{eq:pederson}
   \langle \phi_i | \hat{V}^j | \phi_j \rangle = \langle \phi_i | \hat{V}^i | \phi_j \rangle,
\end{equation}
where $\hat{V}^j$ is the potential arising from the orbital-density-dependent correction felt by orbital $j$, i.e.,
\begin{align}
   \label{eq:pederson2}
   {V}^j(\br)  = \frac{\de{}}{\de{\rho_j(\br)}}\cubra{\alpha \sum_k \Pi_k} \,.
\end{align}
Equation~\eqref{eq:pederson} is known in the literature as the 
Pederson condition.~\cite{pede+84jcp,pede+85jcp,heat+87jcp,pede-lin88jcp,svan+00ijqc,goed-umri97pra,sten-spal08prb}
For the same reason, the $\Lambda$ matrix of Lagrange multipliers appearing in
\begin{equation}
    \hat{H}_{\rm LDA} | \phi_i \rangle +\hat{V}^i | \phi_i \rangle = \sum_j \Lambda_{ji} | \phi_j \rangle,
\end{equation}
is also Hermitean and can then be unitarily diagonalized $\Lambda = U^\dagger \lambda U$.
Besides the $\{ \phi_i \}$ orbitals used to minimize the total energy (also called {\it variational} or
{\it minimizing} orbitals), 
a second set of orbitals can then be introduced by considering the 
eigenvectors of the $\Lambda$ matrix 
\begin{equation}
| \psi_m \rangle= \sum_m | \phi_i \rangle U^\dagger_{im}, 
\end{equation}
which are usually referred to as {\it canonical} 
orbitals,~\cite{pede+84jcp,pede+85jcp,heat+87jcp,pede-lin88jcp,korz+08jcp}
since they are commonly interpreted as the eigenvectors of an effective ODD Hamiltonian (
See Refs.~[\onlinecite{Vydrov2007,sten-spal08prb,ferr+14prb}] for a detailed discussion).

The fact that two sets of orbitals have to be dealt with at the same time
is an important feature of the ODD construction.
In order to illustrate the physical meaning of these orbitals, we first focus on the PZ-SIC functional.
In this case, the variational orbitals may exploit the unitary mixing of Eq.~(\ref{eq:mixingU}) to 
localize (becoming somehow similar to Wannier functions~\cite{wann37pr,marz-vand97prb}),
in order to further lower the total energy. As discussed in Sec.~\ref{sec:OOD_per_atom}, 
this is not always the case, though. 
On the other side, canonical orbitals retain instead the features and the shape of standard electronic-structure 
eigenvectors such as those obtained by KS-DFT or the Hartree-Fock methods.

All the above discussion about variational and canonical orbitals is not limited to PZ-SIC but it also applies to all KC flavors, including KI. For this last functional it is indeed possible, even in the case of full occupations, to define a set of variational orbitals from~\eqn{eq:derivU}, thanks to the definition of KI in \sect{sec:flavors} as the limit,for vanishing weight of the PZ-SIC correction, of the KI$\mhs{L}$ functional.
Without this definition of KI as a limit of KI$\mhs{L}$, any unitary rotation of occupied orbitals could be in principle acceptable. For instance, the canonical orbitals may be chosen, which in the case of KI coincide with the KS eigenstates of the base functional.

\begin{figure}
   \begin{minipage}{.45\linewidth}
   \includegraphics[width=.5\textwidth]{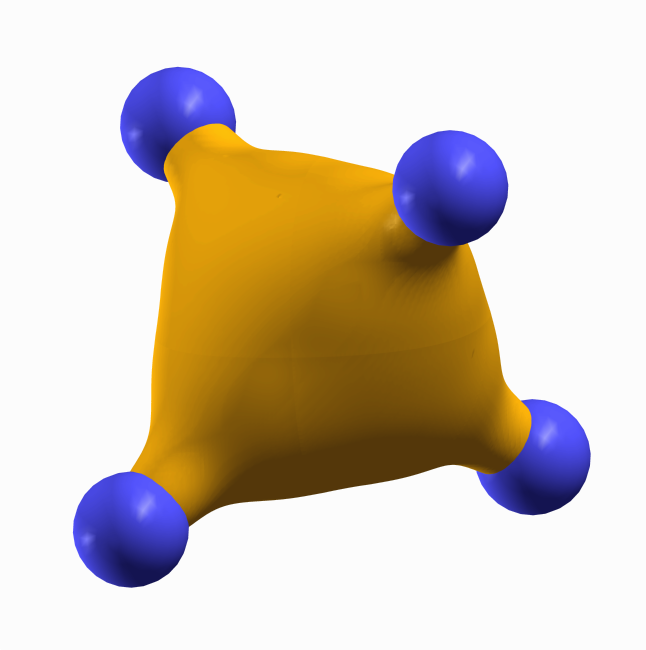}
   \end{minipage}
   \begin{minipage}{.45\linewidth}
   \includegraphics[width=0.5\textwidth]{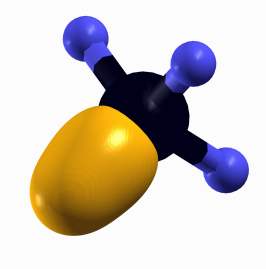}
   \end{minipage}
   \caption{(Color online). An example of canonical (left) and variational (right) orbitals of CH$_4$. 
    Canonical orbitals are eigenvectors of the $\Lambda_{ij}$ matrix and generalizations of KS eigenvalues; 
    variational (minimizing) orbitals are instead those used to define the orbital densities entering in the 
    KC correction.}\label{Fig:canonical_local}
\end{figure}

In \fig{Fig:canonical_local}, we show the canonical and variational orbitals obtained for the
methane molecule by using K [as defined in Sec.~\ref{Sec:KCdef}].
While the former is delocalized over all the molecule, the latter is localized on a single C-H bond.
In closing this section we remark that the total energy of both KC and PZ-SIC functionals
may assume different values when minimized by using an orthogonal rotation instead of
a unitary one in Eq.~(\ref{eq:mixingU}).
As a consequence, the 
minimum of both PZ and KC functionals is found for a manifold of complex-valued
single-particle orbitals~\cite{Klupfel2010,Klupfel2011}, differently to what
happens within KS-DFT, for which the reality (due to time-reversal symmetry) and 
orbital independence of the Hamiltonian results in a purely real
ground-state wave function.
This is further discussed in Sec.~\ref{sec:cmplx_vs_real}.
\begin{figure}
\includegraphics[width=.98\linewidth]{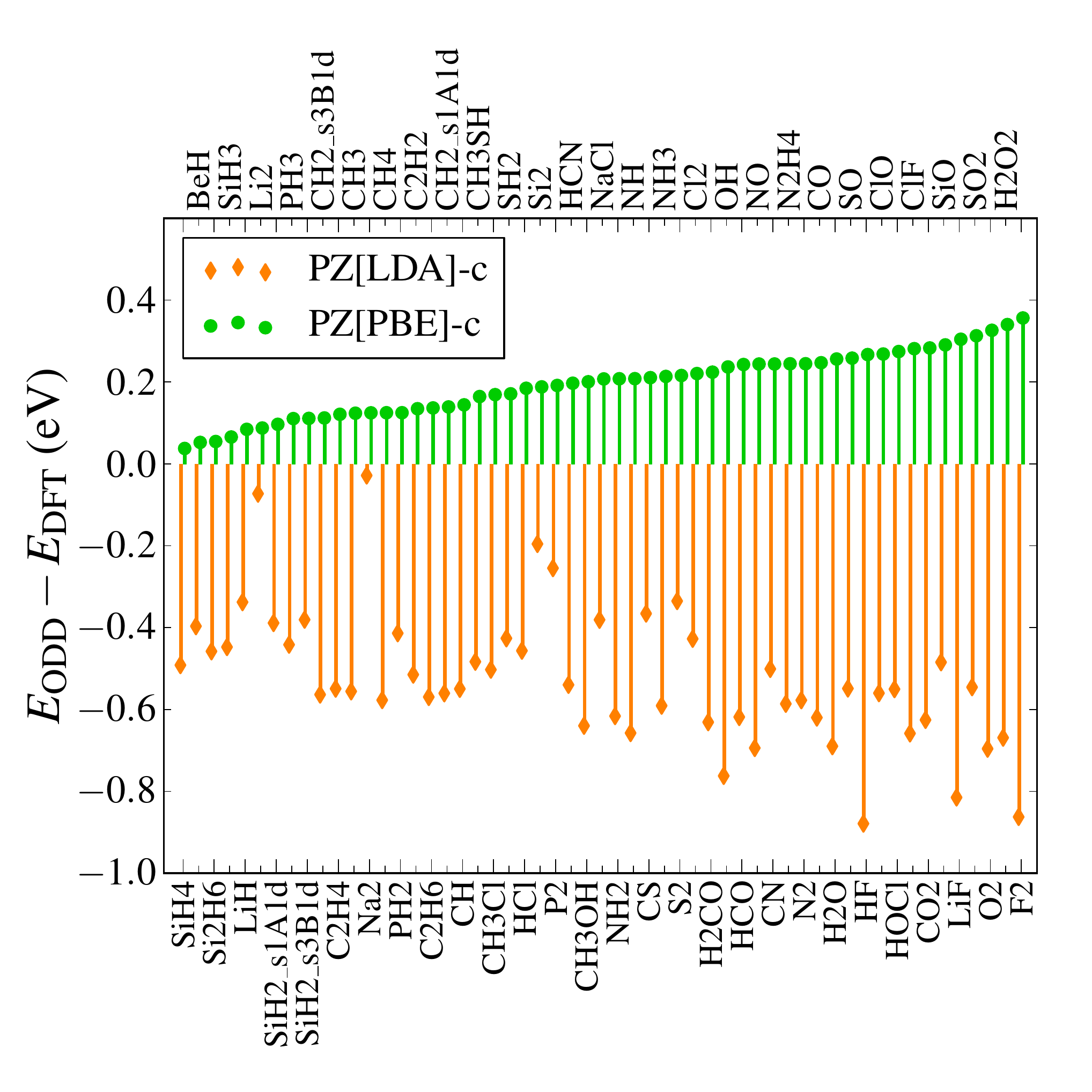}
\caption{(Color online). Orbital-density dependent part of the energy (in eV) per electron for the PZ functional built on top of LDA 
  (orange bars) and PBE (green bars), both are minimized using complex orbitals. 
  Molecules have been arranged in order of increasing PBE energy difference.}\label{Fig:cmplx_pz_eodd}
\end{figure}
\subsection{ODD energy gain per electron}
\label{sec:OOD_per_atom}
\begin{figure}
\includegraphics[width=0.45\textwidth]{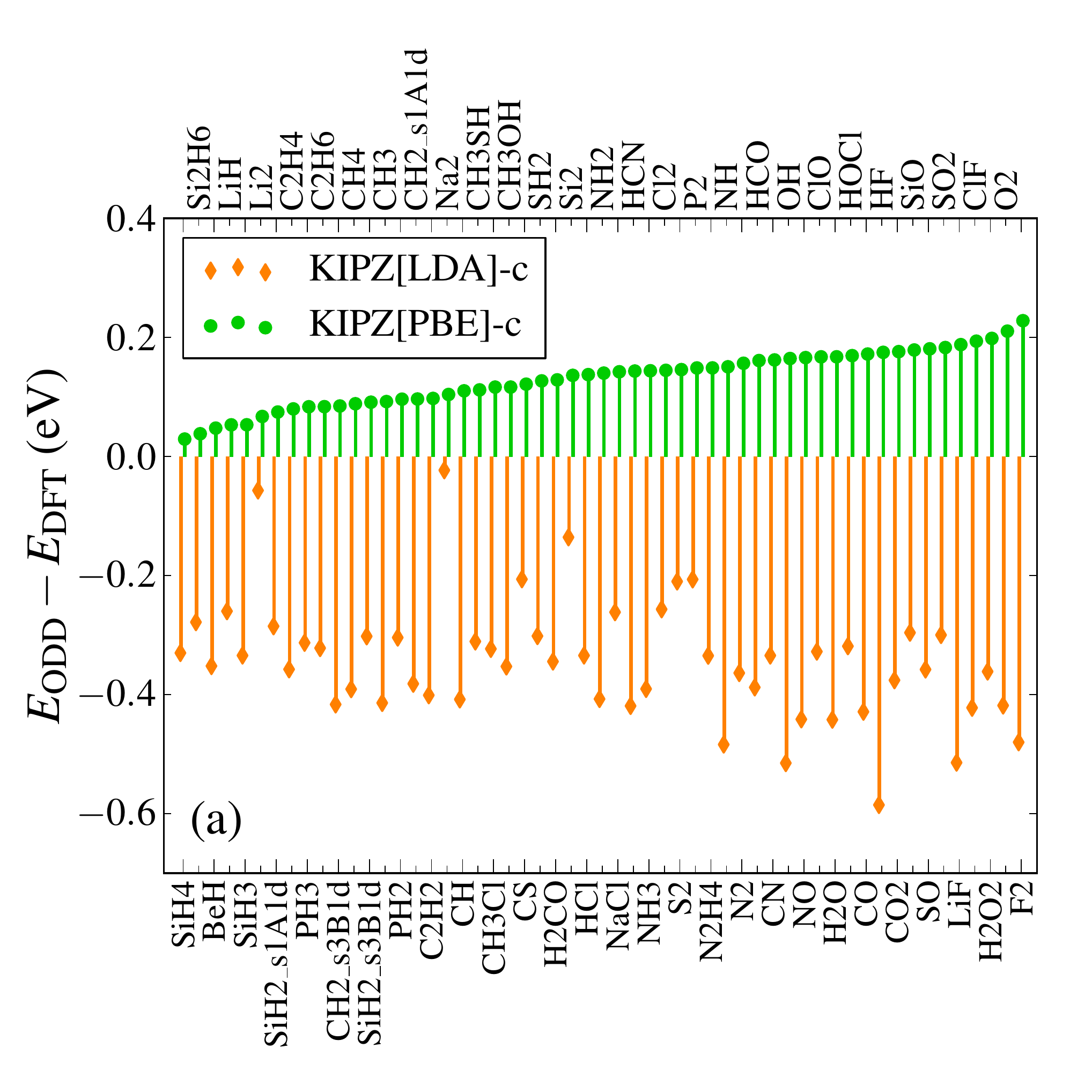}
\includegraphics[width=0.45\textwidth]{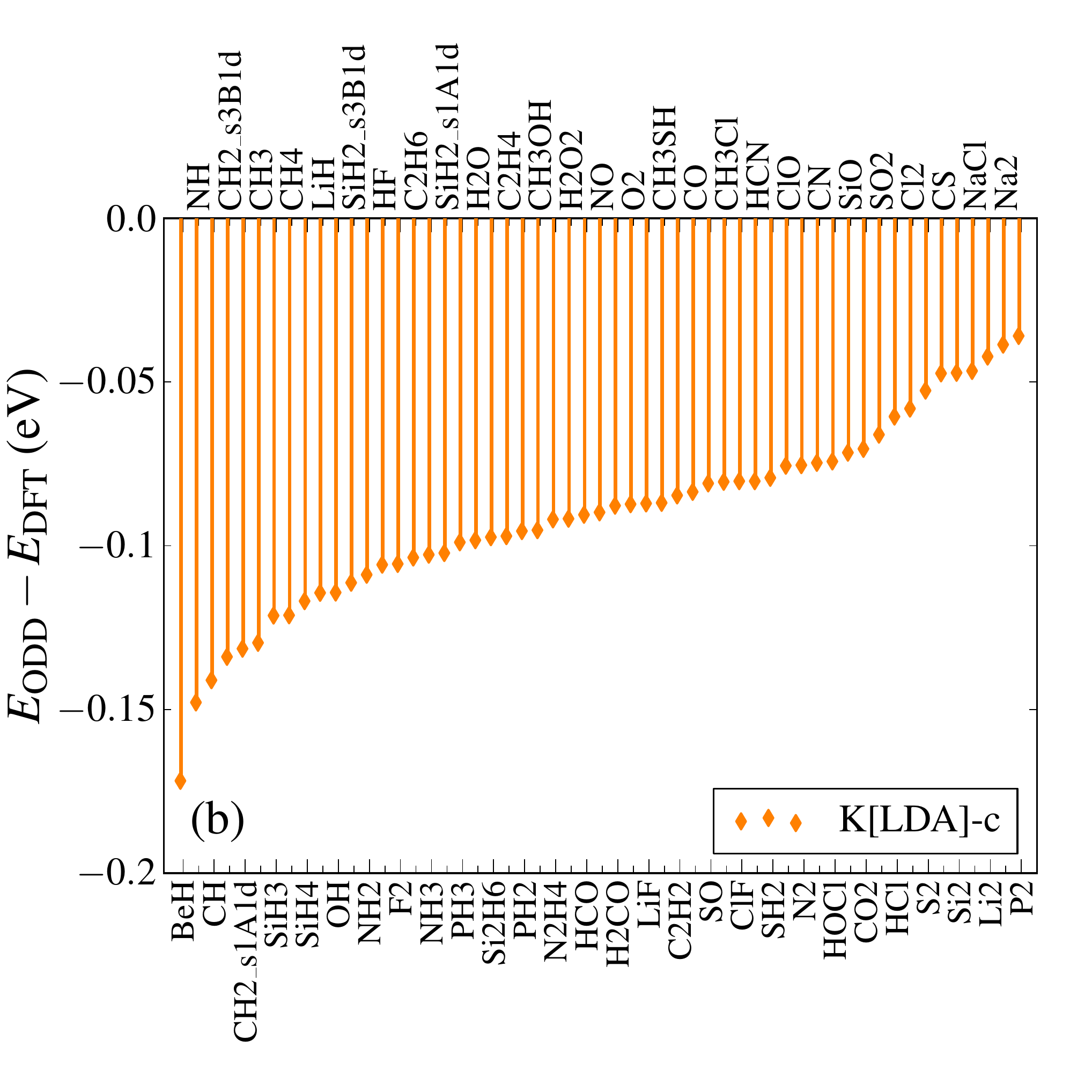}
\caption{(Color online). ODD energy gain (in eV) per electron. (a) upper panel: 
    data for the KIPZ functional built on top of LDA (orange bars) and PBE (green bars).
    (b) Lower panel: data for the K functional on top of LDA.
    All results have been obtained by minimizing the energy using complex orbitals. 
    Molecules have been arranged in order of increasing PBE energy difference.}\label{Fig:cmplx_eodd_nkipznk}
\end{figure}

\begin{figure}
\begin{minipage}{.95\linewidth}
\includegraphics[width=.98\linewidth]{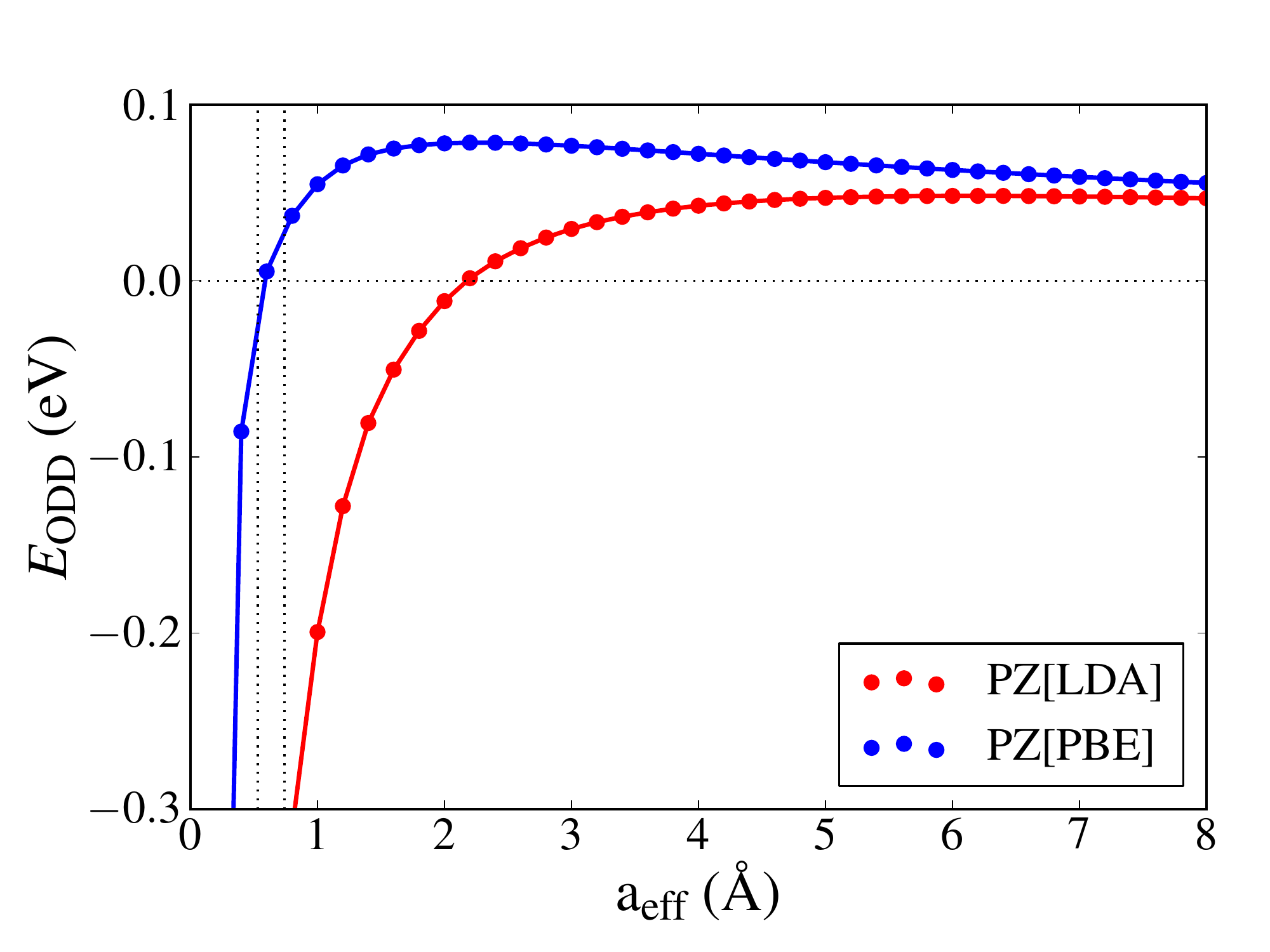}
\caption{(Color online) Plot of PZ orbital-density-dependent correction for a 1$s$ orbital with effective Bohr radius $a_{\rm eff}$. Lines are linear interpolations of a mesh of data. The black line shows the result for LDA, while the blue line for PBE. The two vertical dashed lines mark the values of 0.529~\AA (the Bohr radius) and 0.7 \AA (approximately equal to the bond length of H$_2$). The correction on top of LDA is negative for $a_{\rm eff}\lessapprox 2.5$~\AA, while the correction on top of PBE is positive whenever $a_{\rm eff}\gtrapprox a_{\rm Bohr}$. The black LDA curve is analogous to the one plotted for Gaussian orbitals in the work of K\"orzd\"orfer \etal~\cite{Korzdorfer2011}.}\label{Fig:1s_sic}
\end{minipage}
\end{figure}

Apart from the KI scheme, each KC flavor with a variational energy functional is characterized by a different energy with respect to LDA or PBE. This energy change can be both positive or negative, but our results on the G2-1 set show that the energy change is mostly negative in the case of PBE, while positive in the case of LDA, as shown by Figs.~\ref{Fig:cmplx_pz_eodd} and \ref{Fig:cmplx_eodd_nkipznk}. A justification for this fact in the case of the PZ functional comes from the plot of \fig{Fig:1s_sic}, which shows that, for a hydrogen-1$s$ shaped orbital, the PZ orbital-density-dependent correction for PBE is positive if the orbital has an effective Bohr radius larger than the Bohr radius of the hydrogen atom, while the PZ correction on top of LDA is negative for all radii smaller than approximately 2.5~\AA. Although the PZ correction is applied on molecular rather than atomic orbitals, the above remark tells us that we should expect the PZ correction on top of PBE to be mostly positive, while the one 
on top of LDA to be mostly negative, at least for second and third-row elements, all having Bohr radii with values between 0.5 and 2.5~\AA. 
An argument similar to the above can be applied to the KIPZ functional, which, for screening coefficient $\alpha=1$, has exactly the same energy, for integer particle numbers, as the PZ functional.

The case of the K flavor is different from both PZ and KIPZ. For this scheme, which we use only on top of LDA, the energy change due to the ODD correction appears in any case to be always negative.

\subsection{Complex versus real wave functions}
\label{sec:cmplx_vs_real}
\begin{figure}
\begin{minipage}{.95\linewidth}
\includegraphics[width=.98\linewidth]{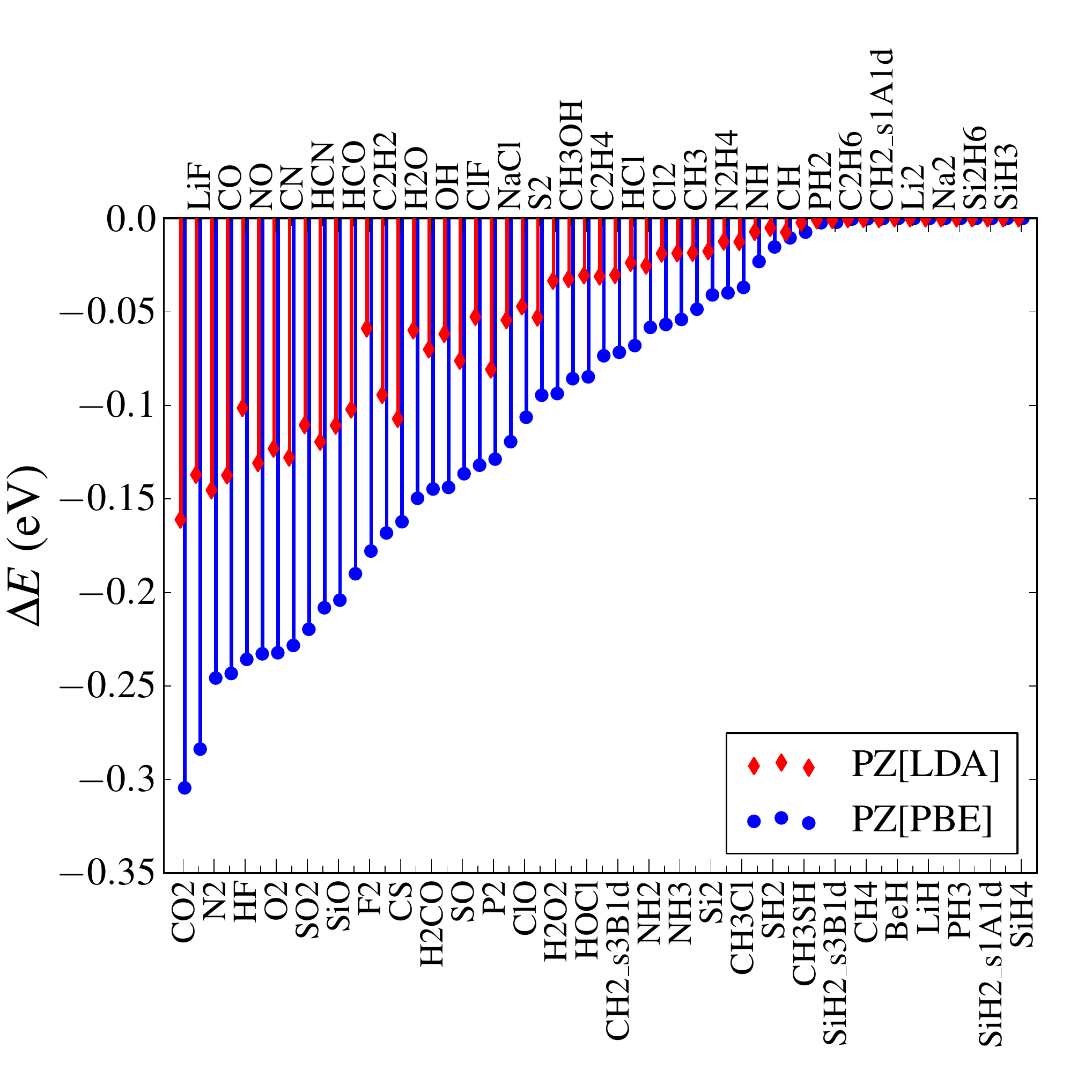}
\end{minipage}
\caption{(Color online). Energy difference per electron (in eV) between the complex-valued and the real-valued wave-function minimum for the PZ functional. The red bars show the result for the LDA functional, while the blue bars show the same for PBE. }\label{Fig:cmplx_egain_eodd_pz}
\end{figure}

There is also a sizable difference between the ODD energy change obtained by minimizing the KC or PZ functional on the Hilbert space of complex wave functions rather than on the smaller set of real wave functions. In 
\fig{Fig:cmplx_egain_eodd_pz}, we show $\Delta E=E_{\rm complex} - E_{\rm real}$ for the PZ functional. This is the functional showing the largest differences between complex and real ground-state energies, and the only one for which the difference is guaranteed to be strictly negative. The same cannot be said in principle for KC functionals for which a difference may exist between screening coefficients for real and complex ground-states, even though, as pointed out in \sect{sec:screening}, this difference appears to be very small.

We can see from \fig{Fig:cmplx_egain_eodd_pz} that in the case of PZ there is a group of molecules for which there appears to be no difference between the energies obtained with real and with complex wave functions. This group contains a class of molecules built out of atoms with purely $s$-type valence electrons, such as Li$_2$, Na$_2$, BeH, LiH, but also other compounds such as PH$_3$, Si$_2$H$_6$, SiH$_3$, SiH$_4$, which are molecules containing $p$-type orbitals with $sp^3$ hybridization. Most oxygen compounds show a fairly large degree of ``complexification'', and in general the presence of double or triple bonds shows a sizable energy gain when minimizing with respect to complex orbitals. 

In~\appendixx{App:complex_explanation}, we show, in a purely atomic and spherically symmetric picture, how the orbitals with a complex $p$-type spherical harmonic as their angular part have a net gain in the PZ Hartree+exchange self-interaction correction energy with respect to orbitals whose angular part is a real $p$-type spherical harmonic.
The above ``complexification'' picture holding for the PZ-SIC functional seems to be valid also for K0 and KIPZ (see Fig.~8 of the Supplemental Material~\cite{Supplemental_material}), while by definition there is no effect of complex orbitals on the energy per electron of KI. This does not imply that the eigenvalue spectrum of KI does not change with the use of complex rather than real orbitals, since the spectrum is determined by the values of the KI scalar potential [see~\eqn{Eq:pot_KI_sc}] which can be different when evaluated on complex orbitals. The same total energy behavior of KI appears in the K functional, where the introduction of complex degrees of freedom does not change the results for the energy per electron.
\subsection{Screening coefficient}\label{sec:screening}
The calculation of the screening coefficient $\alpha$ is crucial in order to include in the KC correction the effects of relaxation of the manifold of single-particle orbitals when the occupation of one of them is changed by a finite amount, so that the single-particle density-matrix of the system is moved away from idempotency.
\begin{figure}
\includegraphics[width=0.45\textwidth]{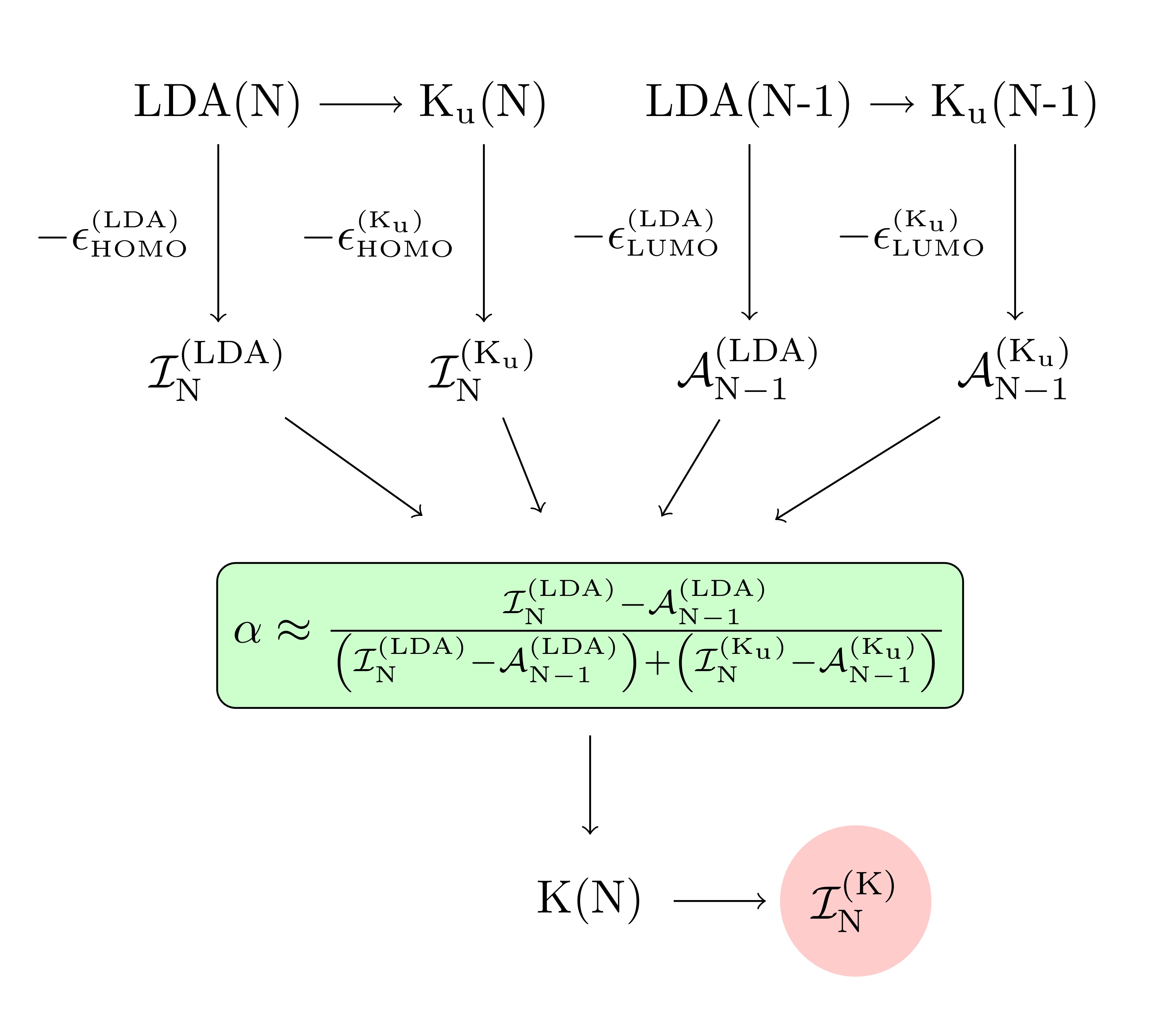}
\caption{Schematic representation of the procedure to obtain the {\it ab initio} estimate of $\alpha$ from calculations on the neutral and singly-ionized molecule. In this work, the first estimate of $\alpha$ is refined one step further with respect to what is shown in this graph (the full secant recursion procedure was presented by Dabo \etal~\cite{Dabo2010}).}\label{Graph:alpha_calc}
\end{figure}

The procedure to compute $\alpha$ for a finite system was devised by Dabo \etal~\cite{Dabo2010}, and it consists in imposing the equality of the values of $\epsilon_{\rm HO}$ (equal to minus the estimate ${\cal I}$ of the ionization energy) of a particular system and $\epsilon_{\rm LU}$ (equal to minus the estimate ${\cal A}$ of the electron affinity) of the same system deprived of an electron.
This condition, while requiring only to perform KC functional minimizations at integer particle number, automatically enforces the piecewise linearity of the energy with respect to fractional changes in the number of particles~\cite{Dabo2010, Dabo2013, psik_koopmans} (see also~\fig{Fig:piecewise}).
It can be imposed exactly in an iterative way using the secant method~\cite{Dabo2010}, using $\epsilon_{\rm HO}(N)-\epsilon_{\rm LU}(N-1)$ [or equivalently ${\cal I}(N)-{\cal A}(N-1)$] as the function of $\alpha$ for which to find a zero within the search interval (0,1).
 
\begin{figure}
\includegraphics[width=.45\textwidth]{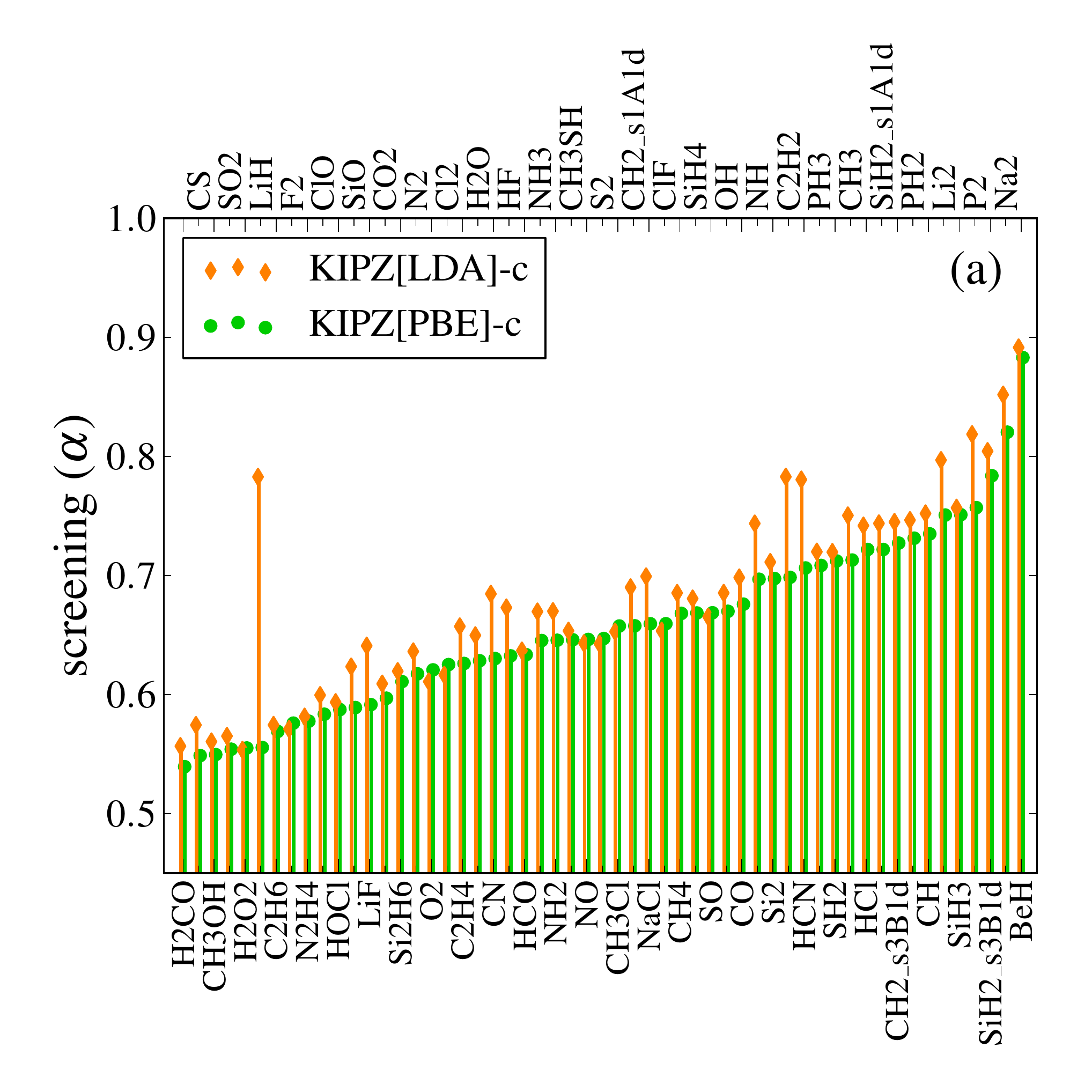}
\includegraphics[width=.45\textwidth]{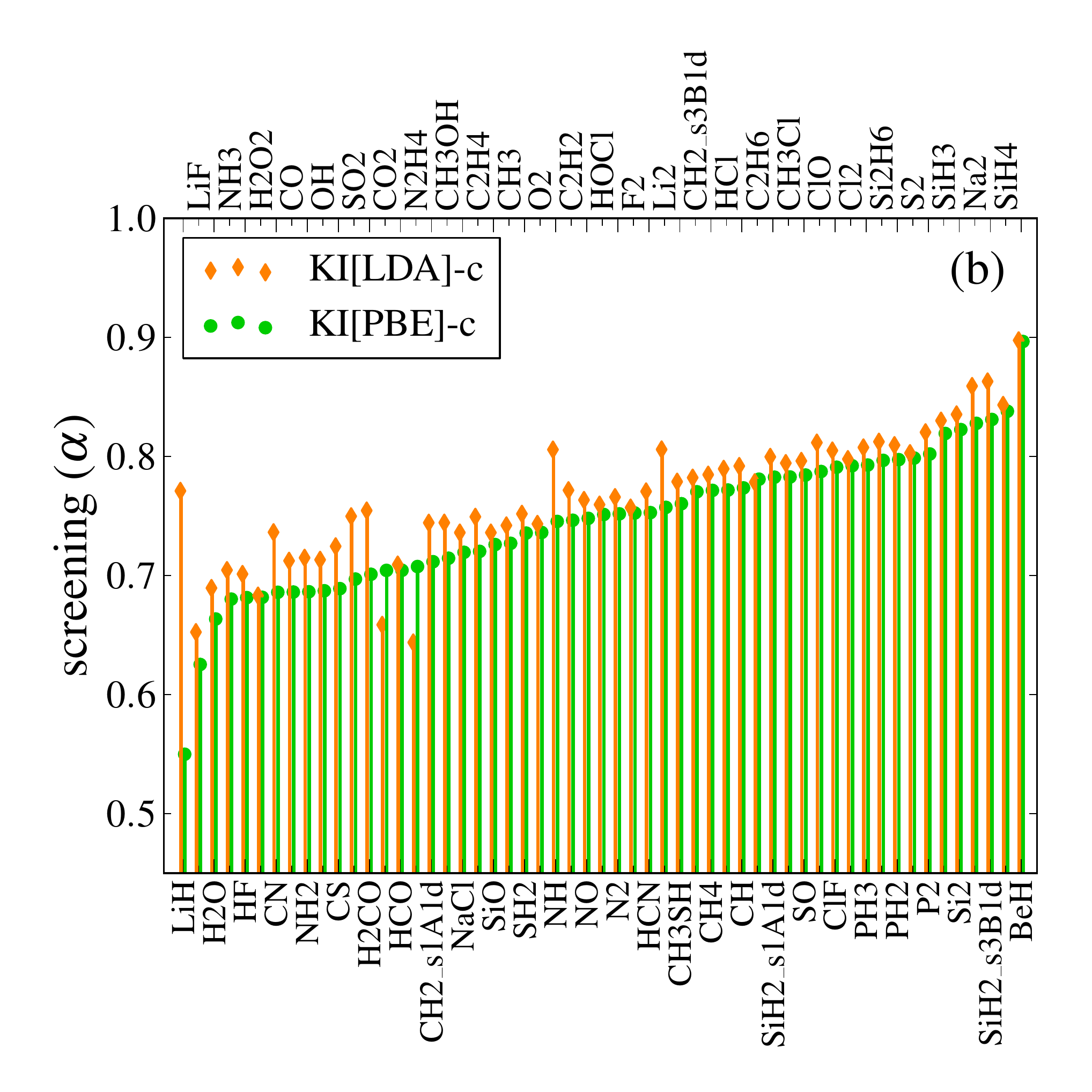}
\caption{(Color online). Screening coefficient $\alpha$ for all G2-1 set molecules. 
      (a) Upper panel: values computed using the KIPZ scheme on top of LDA (orange) and PBE (green bars). 
      (b) Lower panel: values for KI on top of LDA and PBE. All functionals 
      were minimized in the space of complex wave functions. 
      Molecules have been arranged in order of increasing PBE screening 
      coefficients $\alpha$.}\label{Fig:cmplx_alpha_nki_nkipz}
 \end{figure}

In \fig{Graph:alpha_calc}, we sketch the procedure to find the first estimate of $\alpha$, which is already quite accurate in enforcing piecewise linearity. In this work, we further refined this first estimate by performing one extra iteration of the secant method.
In the two panels of Fig.~\ref{Fig:cmplx_alpha_nki_nkipz} we report the values of $\alpha$ computed for the KIPZ and KI functionals, minimized on the space of complex orbitals. For most molecules, there appears to be no dependence on the base functional (LDA or PBE) on top of which the KC correction is applied. Also, when the minimization is performed with real orbitals the values of $\alpha$ show no substantial change with respect to the complex case [compare Fig.~9 of the Supplemental Material~\cite{Supplemental_material} with Fig.~\ref{Fig:cmplx_alpha_nki_nkipz}(a)]. 
The change in functional flavor does instead influence the value of the screening. A general trend that we observe is that the KIPZ $\alpha$ coefficients tend to be smaller than the KI ones, the decrease being particularly large for molecules such as C$_2$H$_6$ and Si$_2$H$_6$.
 
\section{Conclusions}
In this paper, we have assessed the performance of Koopmans-compliant functionals in calculating
ionization energies, geometries, and atomization energies of all molecules in the G2-1 test set, 
showing the accuracy of the method against experimental results.
For ionization energies, we compared the performance of KC approaches with 
that of local and semilocal KS-DFT, PZ-SIC, and many-body perturbation theory (G$_0$W$_0$ and scfGW).
The accuracy of the KC approaches has been found to be as good as or better than that of GW, at a fraction of the computational cost.
While the KC construction always improves on ionization energies,
such accuracy is not automatically transferred to charge localization and total energy differences.
In fact, the K and KI functionals leave the total energies almost or exactly unchanged.
We have linked this issue to the emergence of scalar orbital-dependent potentials (i.e., orbital-dependent energy shifts), and have proposed new flavors for KC functionals (KPZ and KIPZ) 
that aim at canceling these undesired contributions. 
For geometries, the KIPZ functional on top of PBE provides the best estimates 
for bond lengths. For atomization energies, for which PBE performs already very well (as does, identically, KI), the KPZ and KIPZ functionals greatly improve the largely over estimated predictions of LDA, or the results of K and PZ-SIC.
Last, we have investigated some numerical aspects related to the use of ODD functionals and
we have shown that,
as predicted in the case of atomic systems~\cite{Klupfel2011},
the use of complex wavefunctions leads to lower-energy minima, e.g., in compounds containing elements with valence $p$ electrons. This energy gain is sizable in oxygen and halogen 
compounds and in molecules with double or triple bonds. Our results show that complex wave functions 
do not play a significant role for KC calculations, 
while they are important (and generally improve results) when ionization energies are computed with the PZ-SIC scheme (which performs best when applied on top of the PBE functional using complex wavefunctions), or with the KPZ and KIPZ schemes.

Regarding the scaling of the method, in plane-wave implementations
a single conjugate-gradient minimization step has a computational cost which scales as $N$ times the cost of a typical DFT step ($N$ being the number of electrons) in the 
FFT-dominated regime, i.e., $N^2\log(N)$, while it scales exactly as DFT, i.e. $N^3$ in the orthogonalization-dominated regime.

\section*{Acknowledgements}
We thank L. Kronik, E. Kucukbenli, M. Cococcioni and N. Poilvert for useful discussions, G. Pizzi and A. Cepellotti for sharing the AiiDA~\cite{aiida_paper} platform to run and organize multiple computer simulations. We thank the French National Research Agency, Grant No. ANR 12-BS04-0001 PANELS 424 (I.D.) for support. A.F. acknowledges partial support from Italian MIUR through grant FIRB-RBFR08FOAL.
\appendix

\vspace{1cm}
\section{Koopmans-compliant functionals, energy and potential corrections}\label{App:KC}

In this Appendix we provide explicit expressions for the energy and orbital-density-dependent potential corrections of all KC flavors. The results shown are the frozen-orbital corrections: one needs to multiply them by the system-dependent factor $\alpha$ in order to get the screened correction.
In the equations that follow we will use the notation $\rho_{\alpha\sigma}(\br)=f_{\alpha\sigma} n_{\alpha\sigma}(\br)$, $n_{\alpha\sigma}(\br)=|\phi_{\alpha\sigma}(\br)|^2$, indicating therefore with $n_{\alpha\sigma}(\br)$ the occupation-independent part of the orbital-density, and splitting the orbital and the spin indices.

\subsection{The K functional}\label{App:K}

\begin{figure}
\includegraphics[width=0.90\linewidth]{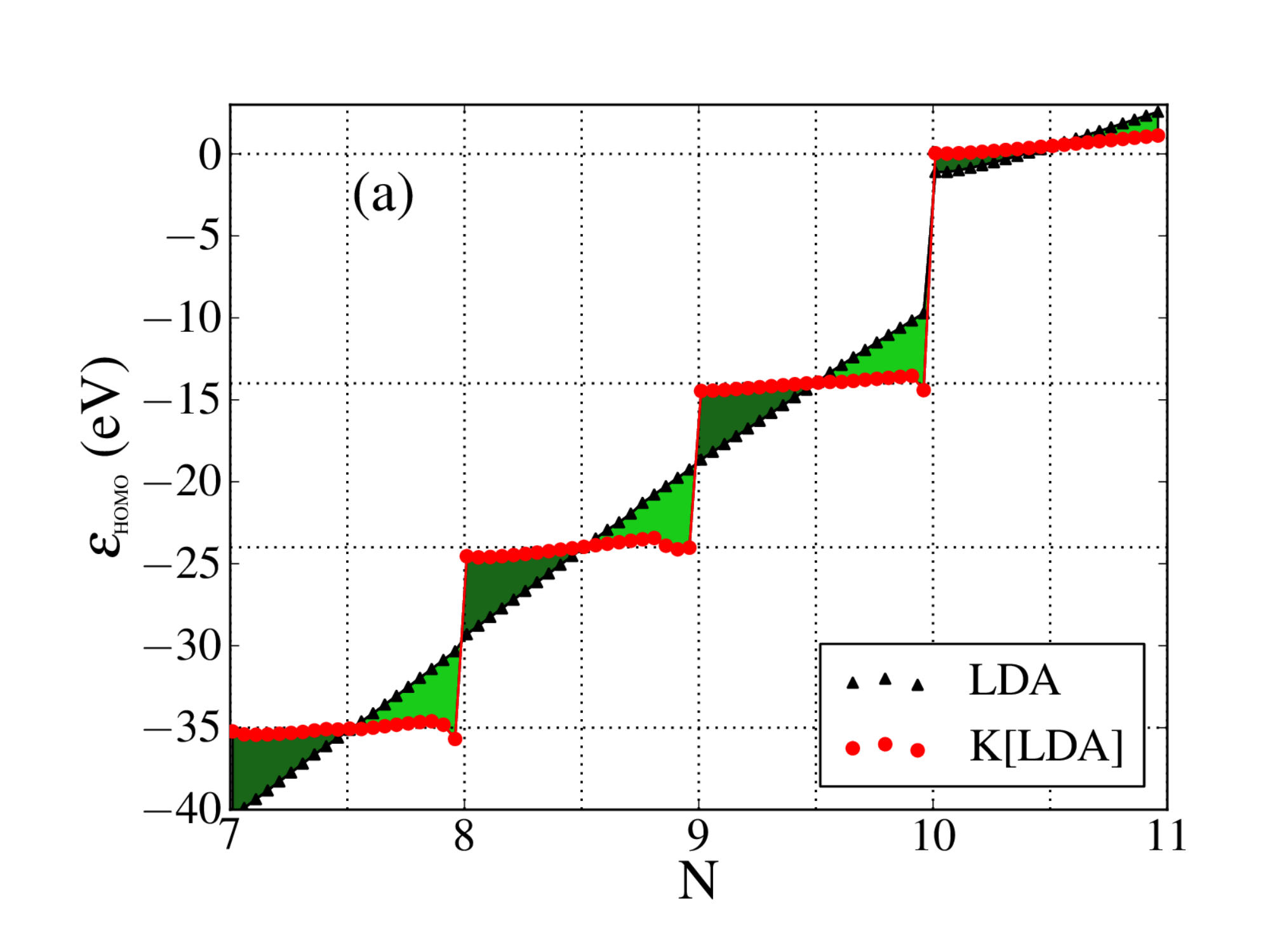}
\caption{(Color online). Eigenvalue of the highest-occupied molecular orbital for the methane molecule with different values of fractional occupation. The area colored in green equals the total energy correction introduced by the K functional on top of LDA through all orbital-density-dependent $\Pi_i$ terms. Bright green areas correspond to negative corrections, while dark green areas are positive corrections. The fact that the area of each bright green ``triangle'' approximately equals the area of the dark green ``triangle'' within the same two integer occupations helps to clarify why the K functional approximately preserves the LDA total energy for integer occupations.}
  \label{Fig:area_piecewise-linearity}
\end{figure}

For the K functional, \eqn{Eq:KC_def} leads to the following energy correction term
\begin{align}\label{Eq:Pi_K}
 \Pi^{\inbra{\rm K}}_{\alpha\sigma} &= -\EHxc[\rho]+\EHxc[\rho-\rho_{\alpha\sigma}]  \nonumber \\
 &+\int \de{\br} \rho_{\alpha\sigma}(\br) {\vHxc}_{\sigma}(\br, [\rho^{\rm ref}_{\alpha\sigma}])\,,
\end{align}
with $\rho^{\rm ref}_{\alpha\sigma}(\br)=\fref n_{\alpha\sigma}(\br)$, $\fref$ being equal to $1/2$.
The derivative of \eqn{Eq:Pi_K} with respect to a change in one orbital density $\rho_{\beta\sigmap}(\br)$ has two contributions, the first coming from the energy change at fixed Hartree-exchange-correlation potential (i.e. constant reference density), and the second including the variation of this potential with respect to density:
\begin{align}\label{Eq:K_pot}
& \robra{\frac{\delta \Pi^{\inbra{\rm K}}_{\alpha\sigma}}{\delta \rho_{\beta\sigmap}(\br)}}_{\rm K} = 
\robra{\frac{\partial \Pi^{\inbra{\rm K}}_{\alpha\sigma}}{\partial \rho_{\beta\sigmap}(\br)}}\Big|_{c.{\vHxc}} + \robra{\frac{\partial \Pi^{\inbra{\rm K}}_{\alpha\sigma}}{\partial \rho_{\beta\sigmap}(\br)}}\Big|_{\delta \vxc}\,.
\end{align}
The two contributions read, respectively:
\begin{align}
\robra{\frac{\partial \Pi^{\inbra{\rm K}}_{\alpha\sigma}}{\partial \rho_{\beta\sigmap}(\br)}}\Big|_{c.{\vHxc}} &\!\!\!\!\!=
-{\vHxc}_{\sigmap}(\br, [\rho])+{\vHxc}_{\sigma}(\br, [\rho^{\rm ref}_{\alpha\sigma}])\delta_{\alpha\beta}\delta_{\sigma\sigmap}\nonumber\\ 
&+{\vHxc}_{\sigmap}(\br, [\rho-\rho_{\alpha\sigma}]) (1-\delta_{\alpha\beta}\delta_{\sigma\sigmap})\label{Eq:K_pot_terms1}\,,\\
\robra{\frac{\partial \Pi^{\inbra{\rm K}}_{\alpha\sigma}}{\partial \rho_{\beta\sigmap}(\br)}}\Big|_{\delta \vHxc} &\!\!\!\!\!= \fref\delta_{\alpha\beta}\delta_{\sigma\sigmap} \int n_{\alpha\sigma}(\brp) \mhs{H}_{\alpha\sigma}(\brp,\br)\de{\brp}+ \nonumber\\
+(1-\delta_{\sigma\sigmap}&\delta_{\alpha\beta}) \int{\kHxc}_{\sigma\sigmap}(\brp,\br,[ \rhoref_{\alpha\sigma}])  \rho_{\alpha\sigma}(\brp)\de{\brp}\,,\label{Eq:K_pot_terms2}
\end{align}
where
\begin{align}
\mhs{H}_{\alpha\sigma}(\brp,\br)& ={\kHxc}_{\sigma\sigma}(\brp,\br, [\rhoref_{\alpha\sigma}])\nonumber\\
&  -\int  {\kHxc}_{\sigma\sigma}(\brp,\br'', [\rhoref_{\alpha\sigma}]) n_{\alpha\sigma}(\br'')\de{\br''}\label{Eq:K_H_term}\,.
\end{align}
Notice the presence, in the above expressions, of the Hartree-exchange-correlation kernel ${\kHxc}_{\sigma\sigmap}$, coming from the variations of the Kohn-Sham potential with respect to orbital densities.
The double-integral on the right-hand side of \eqn{Eq:K_pot_terms2} marks the presence of an $\br$-independent potential term (thus constant in space). Such a ``scalar'' term would have no effect if it belonged to an orbital-independent Kohn-Sham potential. It is the fact that this scalar term has different values for different orbitals that determines its effectiveness in shifting the Kohn-Sham eigenvalues and correcting the value of $\epsilon_{\rm HOMO}$.
As a last remark, we would like to stress that the K total energy correction, although being finite, is typically very small for completely filled orbitals $f_i=1$. To understand that, it is enough to look at \fig{Fig:area_piecewise-linearity}, where the total energy correction introduced by K on top of LDA is represented as the difference between the integrals of the curves for the LDA and the K HOMO eigenvalues.

In the case of KI, it will be easy to see already from its definition how its total energy correction completely vanishes for filled orbitals.
\subsection{The KI functional}\label{App:KI}
The KI energy correction reads as
\begin{align}\label{Eq:Pi_KI}
\Pi^{\inbra{\rm KI}}_{\alpha\sigma} &= -\EHxc[\rho]+\EHxc[\rho-\rho_{\alpha\sigma}] \nonumber\\ &+f_{\alpha\sigma}\robra{-\EHxc[\rho-\rho_{\alpha\sigma}]+\EHxc[\rho-\rho_{\alpha\sigma}+\nalsi]}\,.
\end{align}
As one may realize by setting $f_i=1$ or 0, the correction is identically zero for integer values of the occupations. This implies that the KI functional modifies the LDA energy only within the intervals between integer values of the particle number.
The correction to the LDA potential introduced with this term is
\begin{align}\label{Eq:pot_KI}
&\robra{\frac{\delta \Pi^{\inbra{\rm KI}}_{\alpha\sigma}}{\delta \rho_{\beta\sigmap}(\br)}} = \robra{\frac{\delta \Pi^{\inbra{\rm KI}}_{\alpha\sigma}}{\delta \rho_{\beta\sigmap}(\br)}}\Bigg|_{\rm sc.}+\robra{\frac{\delta \Pi^{\inbra{\rm KI}}_{\alpha\sigma}}{\delta \rho_{\beta\sigmap}(\br)}}\Bigg|_{\rm re.}\,,
\end{align}

where the first term is scalar ($\br$-independent) and is completely diagonal in the orbital index

\begin{align}\label{Eq:pot_KI_sc}
&\robra{\frac{\delta \Pi^{\inbra{\rm KI}}_{\alpha\sigma}}{\delta \rho_{\beta\sigmap}(\br)}}\Bigg|_{\rm sc.}\!\!\!
\!=\Big\{\!\!-\EHxc[\rho-\rho_{\alpha\sigma}]+\EHxc[\rho-\rho_{\alpha\sigma}+\nalsi]+\nonumber \\
&-\int \vHxc_{\sigmap}(\brp,[\rho-\rho_{\alpha\sigma}+\nalsi]) n_{\alpha\sigma}(\brp) \de{\brp}\Big\}\delta_{\alpha\beta}\delta_{\sigma\sigmap}\,,
\end{align}

while the second term is a real-space potential, with both diagonal and off-diagonal terms:

\begin{align}\label{Eq:pot_KI_real}
&\robra{\frac{\delta \Pi^{\inbra{\rm KI}}_{\alpha\sigma}}{\delta \rho_{\beta\sigmap}(\br)}}\Bigg|_{\rm re.}\!\!\!\!\!
\!=\delta_{\alpha\beta}\delta_{\sigma\sigmap}v^{\inbra{\rm KI}}_{\alpha\sigma}(\br)\big|_{\rm d}\!+(1-\delta_{\alpha\beta}\delta_{\sigma\sigmap})v^{\inbra{\rm KI}}_{\alpha\sigma}(\br)\big|_{\rm od}
\end{align}

with:

\begin{align}\label{Eq:pot_KI_real_d}
v^{\inbra{\rm KI}}_{\alpha\sigma}(\br)\big|_{\rm d}&=-{\vHxc}_{\sigmap}(\br, [\rho])\nonumber\\
&+\vHxc_{\sigmap}(\br,[\rho-\rho_{\alpha\sigma}+\nalsi])\,,
\end{align}

and

\begin{align}\label{Eq:pot_KI_real_od}
v^{\inbra{\rm KI}}_{\alpha\sigma}(\br)\big|_{\rm od}&\!=\!(1\!-\!\falsi){\vHxc}_{\sigmap}(\br,[ \rho\!-\!\rho_{\alpha\sigma}])\!-\!{\vHxc}_{\sigmap}(\br, [\rho])\nonumber\\
&\!+\!f_{\alpha\sigma}\vHxc_{\sigmap}(\br,[\rho\!-\!\rho_{\alpha\sigma}\!+\!\nalsi])\,.
\end{align}

It is interesting to note that in case of integer occupations ($f=0$ or 1), only the diagonal terms are non zero, while in the case of a completely filled orbital ($f=1$) the real-space correction vanishes altogether, and only the first term on the right-hand side of \eqn{Eq:pot_KI} survives. The KI potential for a completely filled orbital is indeed entirely scalar since the KI energy correction for that orbital is identically zero. 
This last fact can be easily understood by setting $f_i=1$ and $n_i=\rho_i$ in \eqn{Eq:Pi_KI} (completely filled orbital), or $f_i=0$ and $\rho_i=0$ (completely empty orbital).
The identically zero KI correction for integer occupations introduces an ambiguity on how to define the KI orbital densities for filled orbitals, and that is why we have made the choice, throughout this paper, to define the KI functional as the limit for $\gamma\to 0$ of the KI$\mhs{L}$ functional, whose real-space potential correction is sufficient, even for vanishing $\gamma$, to univocally define orbital densities.
This choice is such that the completely filled ($f=1$) orbitals on which the KI correction is applied become localized around atomic or bond centers, similarly to what happens for all other KC corrections. The localization criterion is that of minimal total PZ correction within the (fixed) LDA single-particle manifold building the LDA ground-state Slater determinant (only a finite $\gamma$ would change the LDA manifold), and leads therefore to orbitals which are localized, even if not maximally localized. One may want to apply the KI correction on maximally localized Wannier functions, and this would lead to results most probably similar to these.

The choice of localized orbitals for KI is of course not the only legitimate choice, although we believe it is the most effective to correct deviations from piecewise linearity. In a recent paper~\cite{krai-kron13prl} [Eq.~(10)], Kraisler and Kronik propose a correction of the LDA energy gap which depends on HOMO and LUMO orbital densities only, and is formally identical to the KI ODD correction computed on these orbitals.
\subsection{KPZ and KIPZ functionals}\label{App:KL_KIL}
As discussed in \sect{Sec:KCdef}, the K$\mhs{L}$ and KI$\mhs{L}$ functionals are a mathematical tool to seamlessly connect KC functionals applied on top of LDA or PBE, such as K and KI, and KC functionals enforcing Koopmans' correction on top of PZ-SIC, i.e., KPZ and KIPZ. We have shown above the expressions for energy and potential corrections for K and KI, and we show here those of KPZ and KIPZ:
\begin{align}
\Pi^{\inbra{\rm KPZ}}_{\alpha\sigma} &= \Pi^{\inbra{\rm K}}_{\alpha\sigma}-\falsi\int \vHxc_{\sigma}(\brp, [\rhoref])\nalsi(\brp)\de{\brp} \label{Eq:Pi_KPZ}\,,\\
\Pi^{\inbra{\rm KIPZ}}_{\alpha\sigma} &= \Pi^{\inbra{\rm KI}}_{\alpha\sigma}-\falsi\EHxc[\nalsi]\,.\label{Eq:Pi_KIPZ}
\end{align}
Equations~\eqref{Eq:Pi_KPZ} and~\eqref{Eq:Pi_KIPZ} give rise to the following potential corrections:
\begin{align}
\frac{\delta \Pi^{\inbra{\rm KPZ}}_{\alpha\sigma}}{\delta \rho_{\beta\sigmap}(\br)} &= \frac{\delta \Pi^{\inbra{\rm K}}_{\alpha\sigma}}{\delta \rho_{\beta\sigmap}(\br)}-\Big\{\vHxc_{\sigma}(\br,[ \rhoref])\nonumber \\
& +\fref \int \nalsi(\brp)\mhs{H}_{\alpha\sigma}(\brp,\br)\de{\brp}\Big\}\delta_{\alpha\beta}\delta_{\sigma\sigmap}\label{Eq:pot_KPZ}\,,\\
\frac{\delta \Pi^{\inbra{\rm KIPZ}}_{\alpha\sigma}}{\delta \rho_{\beta\sigmap}(\br)} & = \frac{\delta \Pi^{\inbra{\rm KI}}_{\alpha\sigma}}{\delta \rho_{\beta\sigmap}(\br)}- \Big\{\EHxc[\nalsi]+\vHxc_{\sigma}(\br, [\nalsi])\nonumber\\
&-\int \vHxc_{\sigma}(\brp, [\nalsi])n_{\alpha\sigma}(\brp)\de{\brp}\Big\}\delta_{\alpha\beta}\delta_{\sigma\sigmap}\label{Eq:pot_KIPZ}\,.
\end{align}
It is not difficult to verify that the orbital-density-dependent Hartree energies and potentials do not contribute to the scalar terms of KPZ and KIPZ potentials, contrary to what happens in K and KI, for which finite scalar Hartree potential terms are present. The scalar terms stemming from the exchange-correlation energy are removed only up to second order in the expansion with respect to orbital densities $n_{\alpha\sigma}(\br)$.
One can also easily check that for a system of one electron, KPZ and KIPZ reduce both to PZ-SIC, which is the exact functional for a single electron or for a fractional charge smaller than one. We believe that this feature of being exact in the one-electron limit, while correctly enforcing the piecewise linearity of the energy for any number of electrons, is the reason why these two orbital-density-dependent functionals are not only outperforming PZ-SIC, but also other KC functionals in predicting ionization energies, geometric and energetic properties of molecules.
\subsection{The K0 functional}\label{App:K0}
The K0 non variational functional, introduced for the first time in Ref.~[\onlinecite{Dabo2010}], while having the same expression of \eqn{Eq:Pi_K} for its energy correction, has a potential correction which can be extracted from Eqs.~\eqref{Eq:K_pot}, \eqref{Eq:K_pot_terms1}, and \eqref{Eq:K_pot_terms2} by keeping only the derivative at fixed $\vHxc$, and discarding the cross-orbital terms, so that we get
\begin{align}\label{Eq:K0_pot}
\robra{\frac{\delta \Pi^{\inbra{\rm K}}_{\alpha\sigma}}{\delta \rho_{\beta\sigmap}(\br)}}_{\rm K0} &\!\!\!\!\!=
\sqbra{-{\vHxc}_{\sigmap}(\br, [\rho])
+{\vHxc}_{\sigma}(\br, [\rho^{\rm ref}_{\alpha\sigma}])}\delta_{\alpha\beta}\delta_{\sigma\sigmap}\,.
\end{align}
\section{Heuristic explanation for the complexification of PZ and KIPZ minimizing orbitals}\label{App:complex_explanation}

In this section, we explain the origin of the complexification of the orbitals which minimize the Perdew-Zunger ODD functional (and the KC functional flavors containing a PZ-type term in the potential, such as KPZ and KIPZ).
In order to do this, we assume that most complexification effects can be explained by looking at the Hartree and exchange terms of the PZ and KC corrections only. We will comment on the role of correlations later.
With this assumption, we test the amount of PZ self-interaction energy which can be gained by transforming a hydrogenic orbital with a given radial part and a real spherical harmonic in the angular part, to another orbital in which the real harmonic is turned into a complex harmonic. Taking a $p$ orbital as an example, we can see that the ratio between the Hartree energies computed on the real ($p_z$) and complex ($p_\pm$) orbital is
\begin{align}
\Theta_{\rm H} = \frac{\robra{\EH}_\pm}{\robra{\EH}_z} \approx 0.94\,,
\end{align}
while the ratio between the two exchange energies (taking the Slater expression for exchange) is
\begin{align}
\Theta_{\rm x}=\frac{\robra{\Ex}_\pm}{\robra{\Ex}_z} = \frac{-\int \sqbra{|Y_{1\pm}(\Omega)|^2}^{4/3} d\Omega}{-\int \sqbra{|Y_{10}(\Omega)|^2}^{4/3} d\Omega} \approx 0.89\,.
\end{align}
From the above values for $\Theta_{\rm H}$ and $\Theta_{\rm x}$, it can be easily shown that the following inequality (remembering that Hartree energy is always positive, while the exchange energy is always negative) holds:
\begin{align}
\Pi^{\rm PZ}_\pm &= -|\robra{\EH}_\pm| + |\robra{\Ex}_\pm|\nonumber\\
&=- \Theta_{\rm H}|\robra{\EH}_z| + \Theta_{\rm x}|\robra{\Ex}_z| < \Pi^{\rm PZ}_{z}\,,
\end{align}
every time we have
\begin{align}
\xi_{\rm x}=\frac{|\robra{\Ex}_z|}{|\robra{\EH}_z|} > \bar{\xi}_{\rm x} = 0.55\,.
\end{align}
where $\bar{\xi}_{\rm x}$ is the critical percentage of exchange energy with respect to Hartree energy which will drive a transition to a complex minimizing orbital.

Now, it is well known that the LDA functional under estimates (in absolute value) the exchange energy of an inhomogeneous system, with a typical relative error of around 10\%. It is also known that this under estimation is partly compensated by the over estimation of the correlation energy. While, ideally, for a one-electron system, the exchange energy should exactly cancel the Hartree energy ($\xi_{\rm x}=1$), in the case of the LDA functional computed on a single-electron orbital density, we can safely suppose that $\xi_{\rm x}$ will not be much smaller than 0.9. If we add the correlation energy to exchange, provided that the corresponding ratio $\Theta_{\rm xc}$ does not deviate from the value of $\Theta_{\rm x}$, we will find $\xi_{\rm xc}$ to be even larger, while $\bar{\xi}_{\rm xc}$ will be very close to $\bar{\xi}_{\rm x}$.

We therefore conclude that the complexification of the minimizing orbitals in an ODD density-functional minimization with PZ-type corrections is mainly due to the fact that a complex orbital is characterized by a larger self-exchange energy than a real orbital. The self-Hartree energy loss in going from a real to a complex orbital is too small to prevent the complexification to happen. This statement is valid for LDA, but our results suggest (see \fig{Fig:cmplx_egain_eodd_pz}) that a similar effect might be present in PBE, having possibly a stronger drive to complexification (due to the PBE exchange enhancement factor~\cite{Perdew1996}) than within LDA.
Our conclusion is based on calculations on $p$-type hydrogenic orbitals, and as can be seen from all equations of this section, no assumption has to be made on the radial part of the orbitals. No complexification is expected for $s$-type orbitals, for which the angular part is trivial. We do not dwell here upon the case of $d$-type orbitals, which are not present in the molecules discussed in this paper, but for which we expect that conclusions qualitatively similar to those for $p$-type orbitals can be drawn.

\bibliographystyle{aip}
\bibliography{papers}

\begin{thebibliography}{10}

\bibitem{Hohenberg1964}
P.~Hohenberg and W.~Kohn,
\newblock Phys. Rev. {\bf 136}, B864 (1964).

\bibitem{Kohn1965}
W.~Kohn and L.~J. Sham,
\newblock Phys. Rev. {\bf 140}, A1133 (1965).

\bibitem{perd-levy83prl}
J.~Perdew and M.~Levy,
\newblock Phys. Rev. Lett. {\bf 51}, 1884 (1983).

\bibitem{perd-levy97prb}
J.~Perdew and M.~Levy,
\newblock Phys. Rev. B {\bf 56}, 16021 (1997).

\bibitem{Chong2002}
D.~P. Chong, O.~V. Gritsenko, and E.~J. Baerends,
\newblock The Journal of Chemical Physics {\bf 116}, 1760 (2002).

\bibitem{Perdew1981}
J.~P. Perdew and A.~Zunger,
\newblock Phys. Rev. B {\bf 23}, 5048 (1981).

\bibitem{Perdew1996}
J.~P. Perdew, K.~Burke, and M.~Ernzerhof,
\newblock Phys. Rev. Lett. {\bf 77}, 3865 (1996).

\bibitem{cohe+08sci}
A.~Cohen, P.~Mori-Sanchez, and W.~Yang,
\newblock Science {\bf 321}, 792 (2008).

\bibitem{Dabo2009}
I.~Dabo, M.~Cococcioni, and N.~Marzari,
\newblock arXiv:0901.2637v1  (2009).

\bibitem{Dabo2010}
I.~Dabo, A.~Ferretti, N.~Poilvert, Y.~Li, N.~Marzari, and M.~Cococcioni,
\newblock Phys. Rev. B {\bf 82}, 115121 (2010).

\bibitem{krai-kron13prl}
E.~Kraisler and L.~Kronik,
\newblock Phys. Rev. Lett. {\bf 110}, 126403 (2013).

\bibitem{perd+82prl}
J.~P. Perdew, R.~G. Parr, M.~Levy, and J.~L. Balduz,
\newblock Phys. Rev. Lett. {\bf 49}, 1691 (1982).

\bibitem{Sham1983}
L.~J. Sham and M.~Schl\"uter,
\newblock Phys. Rev. Lett. {\bf 51}, 1888 (1983).

\bibitem{coco-degi05prb}
M.~Cococcioni and S.~D. Gironcoli,
\newblock Phys. Rev. B {\bf 71}, 035105 (2005).

\bibitem{kuli+06prl}
H.~Kulik, M.~Cococcioni, D.~Scherlis, and N.~Marzari,
\newblock Phys. Rev. Lett. {\bf 97}, 103001 (2006).

\bibitem{cococcioni_thesis}
M.~Cococcioni,
\newblock {\em A LDA+U study of selected iron compounds},
\newblock PhD thesis, International School for Advanced Studies (SISSA/ISAS),
  2002.

\bibitem{ruzs+06jcp}
A.~Ruzsinszky, J.~Perdew, G.~Csonka, O.~Vydrov, and G.~Scuseria,
\newblock J. Chem. Phys. {\bf 125}, 194112 (2006).

\bibitem{mori+06jcp}
P.~Mori-S{\'a}nchez, A.~Cohen, and W.~Yang,
\newblock J. Chem. Phys. {\bf 125}, 201102 (2006).

\bibitem{Dabo2013}
I.~Dabo, A.~Ferretti, C.-H. Park, N.~Poilvert, Y.~Li, M.~Cococcioni, and
  N.~Marzari,
\newblock Phys. Chem. Chem. Phys. {\bf 15}, 685 (2013).

\bibitem{lany-zung10prb}
S.~Lany and A.~Zunger,
\newblock Phys. Rev. B {\bf 81}, 205209 (2010).

\bibitem{zhen+11prl}
X.~Zheng, A.~Cohen, P.~Mori-S{\'a}nchez, X.~Hu, and W.~Yang,
\newblock Phys. Rev. Lett. {\bf 107}, 026403 (2011).

\bibitem{refa+11prb}
S.~Refaely-Abramson, R.~Baer, and L.~Kronik,
\newblock Phys. Rev. B {\bf 84}, 075144 (2011).

\bibitem{refa+12prl}
S.~Refaely-Abramson, S.~Sharifzadeh, N.~Govind, J.~Autschbach, J.~Neaton,
  R.~Baer, and L.~Kronik,
\newblock Phys. Rev. Lett. {\bf 109}, 226405 (2012).

\bibitem{kronik_hybrid}
I.~Tamblyn, S.~Refaely-Abramson, J.~B. Neaton, and L.~Kronik,
\newblock J. Phys. Chem. Lett. {\bf 5}, 2734 (2014).

\bibitem{psik_koopmans}
I.~Dabo, A.~Ferretti, G.~Borghi, N.~L. Nguyen, N.~Poilvert, C.~H. Park,
  M.~Cococcioni, and N.~Marzari,
\newblock Psi-K Newsletter {\bf 119} (2013).

\bibitem{Slater1974}
J.~C. Slater,
\newblock {\em Quantum Theory of Molecules and Solids: The Self-Consistent
  Field for Molecules and Solids}, volume~4 of {\em International Series in
  Pure and Applied Physics},
\newblock McGraw-Hill, New York, 1974.

\bibitem{Rostgaard2010}
C.~Rostgaard, K.~W. Jacobsen, and K.~S. Thygesen,
\newblock Phys. Rev. B {\bf 81}, 085103 (2010).

\bibitem{Kraisler_Makov}
E.~Kraisler, G.~Makov, N.~Argaman, and I.~Kelson,
\newblock Phys. Rev. A {\bf 80}, 032115 (2009).

\bibitem{Janak1978}
J.~F. Janak,
\newblock Phys. Rev. B {\bf 18}, 7165 (1978).

\bibitem{Giesbertz_Baerends}
K.~J.~H. Giesbertz and E.~J. Baerends,
\newblock The Journal of Chemical Physics {\bf 132},  (2010).

\bibitem{Vydrov2007}
O.~A. Vydrov, G.~E. Scuseria, and J.~P. Perdew,
\newblock The Journal of Chemical Physics {\bf 126}, 154109 (2007).

\bibitem{nguyen_photoemission}
N.~L. Nguyen, G.~Borghi, A.~Ferretti, I.~Dabo, and N.~Marzari,
\newblock to be submitted .

\bibitem{perd+07pra}
J.~Perdew, A.~Ruzsinszky, G.~Csonka, O.~Vydrov, G.~Scuseria, V.~Staroverov, and
  J.~Tao,
\newblock Phys. Rev. A {\bf 76}, 040501 (2007).

\bibitem{pede+84jcp}
M.~Pederson, R.~Heaton, and C.~Lin,
\newblock J. Chem. Phys. {\bf 80}, 1972 (1984).

\bibitem{sten-spal08prb}
M.~Stengel and N.~Spaldin,
\newblock Phys. Rev. B {\bf 77}, 155106 (2008).

\bibitem{Almbladh1985}
C.-O. Almbladh and U.~von Barth,
\newblock Phys. Rev. B {\bf 31}, 3231 (1985).

\bibitem{Supplemental_material}
See Supplemental Material at http: for extra figures and tables showing results
  and supporting the discussions in the paper.

\bibitem{Note1}
It is possible in principle to compute K and KPZ results on top of PBE,
  provided the PBE exchange-correlation kernel can be computed. The calculation
  of the PBE kernel is delicate, and we found that in our code it was a source
  of numerical instability.

\bibitem{G2_web_data}
NIST web data for geometries of molecules in the G2 set,
  \url{http://www.cse.anl.gov/OldCHMwebsiteContent/compmat/G2-97.htm}.

\bibitem{curtiss_reference_g2}
L.~A. Curtiss, K.~Raghavachari, P.~C. Redfern, and J.~A. Pople,
\newblock The Journal of Chemical Physics {\bf 106}, 1063 (1997).

\bibitem{staroverov_g2_reference}
V.~N. Staroverov, G.~E. Scuseria, J.~Tao, and J.~P. Perdew,
\newblock The Journal of Chemical Physics {\bf 119}, 12129 (2003).

\bibitem{Li2011}
Y.~Li and I.~Dabo,
\newblock Phys. Rev. B {\bf 84}, 155127 (2011).

\bibitem{Klupfel2011}
S.~Kl\"upfel, P.~Kl\"upfel, and H.~J\'onsson,
\newblock Phys. Rev. A {\bf 84}, 050501 (2011).

\bibitem{Hofmann2012}
D.~Hofmann, S.~Kl\"upfel, P.~Kl\"upfel, and S.~K\"ummel,
\newblock Phys. Rev. A {\bf 85}, 062514 (2012).

\bibitem{Klupfel2012}
S.~Kl\"upfel, P.~Kl\"pfel, and H.~J\'onsson,
\newblock The Journal of Chemical Physics {\bf 137}, 124102 (2012).

\bibitem{pede+85jcp}
M.~Pederson, R.~Heaton, and C.~Lin,
\newblock Journal of Chemical Physics {\bf 82}, 2688 (1985).

\bibitem{heat+87jcp}
R.~Heaton, M.~Pederson, and C.~Lin,
\newblock Journal of Chemical Physics {\bf 86}, 258 (1987).

\bibitem{pede-lin88jcp}
M.~Pederson and C.~Lin,
\newblock Journal of Chemical Physics {\bf 88}, 1807 (1988).

\bibitem{svan+00ijqc}
A.~Svane, W.~Temmerman, Z.~Szotek, J.~Laegsgaard, and H.~Winter,
\newblock Int. J. Quantum Chem. {\bf 77}, 799 (2000).

\bibitem{goed-umri97pra}
S.~Goedecker and C.~Umrigar,
\newblock Phys. Rev. A {\bf 55}, 1765 (1997).

\bibitem{korz+08jcp}
T.~K{\"o}rzd{\"o}rfer, S.~K{\"u}mmel, and M.~Mundt,
\newblock J. Chem. Phys. {\bf 129}, 014110 (2008).

\bibitem{ferr+14prb}
A.~Ferretti, I.~Dabo, M.~Cococcioni, and N.~Marzari,
\newblock Phys. Rev. B {\bf 89}, 195134 (2014).

\bibitem{wann37pr}
G.~H. Wannier,
\newblock Physical Review {\bf 52}, 191 (1937).

\bibitem{marz-vand97prb}
N.~Marzari and D.~Vanderbilt,
\newblock Phys. Rev. B {\bf 56}, 12847 (1997).

\bibitem{Klupfel2010}
P.~Kl\"upfel, S.~Kl\"upfel, and H.~Jonsson,
\newblock Private publishing  (2010).

\bibitem{Korzdorfer2011}
T.~K\"orzd\"orfer,
\newblock Journal of Chemical Physics {\bf 134}, 094111 (2011).

\bibitem{aiida_paper}
G.~Pizzi, A.~Cepellotti, R.~Sabatini, N.~Marzari, and B.~Kozinsky,
\newblock to be submitted ,
\newblock \url{http://www.aiida.net}.

\end{thebibliography}

\end{document}